# Mechanisms of THz Radiation Generation in Multi-Color Laser-Plasma Interactions: A Review Across Diverse Media

A. A. Molavi Choobini[1]*, S. S. Ghaffari-Oskooei[2], F. M. Aghamir[1], M. Shahmansouri[2]*
[1]*Dept. of Physics, University of Tehran, Tehran 14399-55961, Iran.*
[2]*Department of Atomic and Molecular Physics, Faculty of Physics, Alzahra University, Tehran, Iran.*

***Corresponding authors**: aa.molavich@ut.ac.ir, m.shahmansouri@alzahra.ac.ir

## Abstract:

The exploration of Terahertz (THz) waves has captivated researchers across diverse scientific disciplines such as physics, spectroscopy, chemistry, biology, and engineering, driven by the myriad applications these waves offer. Within this expansive landscape, the development of efficient and reliable THz sources stands as a paramount objective. In the pursuit of this goal, a multitude of approaches have been undertaken, with a notable contender emerging in the form of laser-induced plasma. Harnessing the advancements in ultrafast pulses, laser-induced plasma has proven to be a promising tool for generating THz waves. Its appeal lies in the robust attributes of a high power threshold, intense THz signal, and a broadband THz spectrum. This paper delves into a comprehensive review of the physics and progress underlying THz generation from laser-induced plasmas, exploring scenarios where plasmas are induced in gases, liquids, and solids. The interactions between lasers and plasmas involve complex physical processes, resulting in a variety of laser plasma scenarios for THz generation. In this review, the focus is specifically placed on classifying THz generation based on different physical mechanisms and also examines the characteristics of the emitted THz waves. By categorizing the processes, a deeper understanding of the underlying principles can be attained.

## Contents

## 1 Introduction

Over the past decades, the terahertz (THz) region of the electromagnetic spectrum—situated between the microwave and infrared bands (Fig. 1)—has emerged as a major focus of research. In various units, radiation at 1 THz corresponds to a period of 1 ps, a wavelength of 300 μm, a wavenumber of 33 $cm^{-1}$, a photon energy of 4.1 meV, and an equivalent temperature of 47.6 K [1, 2]. The photon energies are comparable to typical Fermi levels, while the associated electric and magnetic field strengths can reach or exceed the MV/cm and Tesla ranges, respectively. This spectral range enables unique interactions between light and matter, resonantly coupling to a variety of fundamental excitations of ions, electrons, and spins in all phases of matter. Intense THz fields at specific frequencies can coherently drive lattice resonances, modify electronic structures, and induce novel states [3]. Owing to these capabilities, the so-called "THz gap" is rapidly closing, giving rise to the concept of a "THz bridge" [4]. Today, THz radiation finds diverse applications in medicine, microelectronics, agriculture, forensic science, and beyond, supported by its ability to reveal distinct spectroscopic "fingerprints" of materials and phenomena.

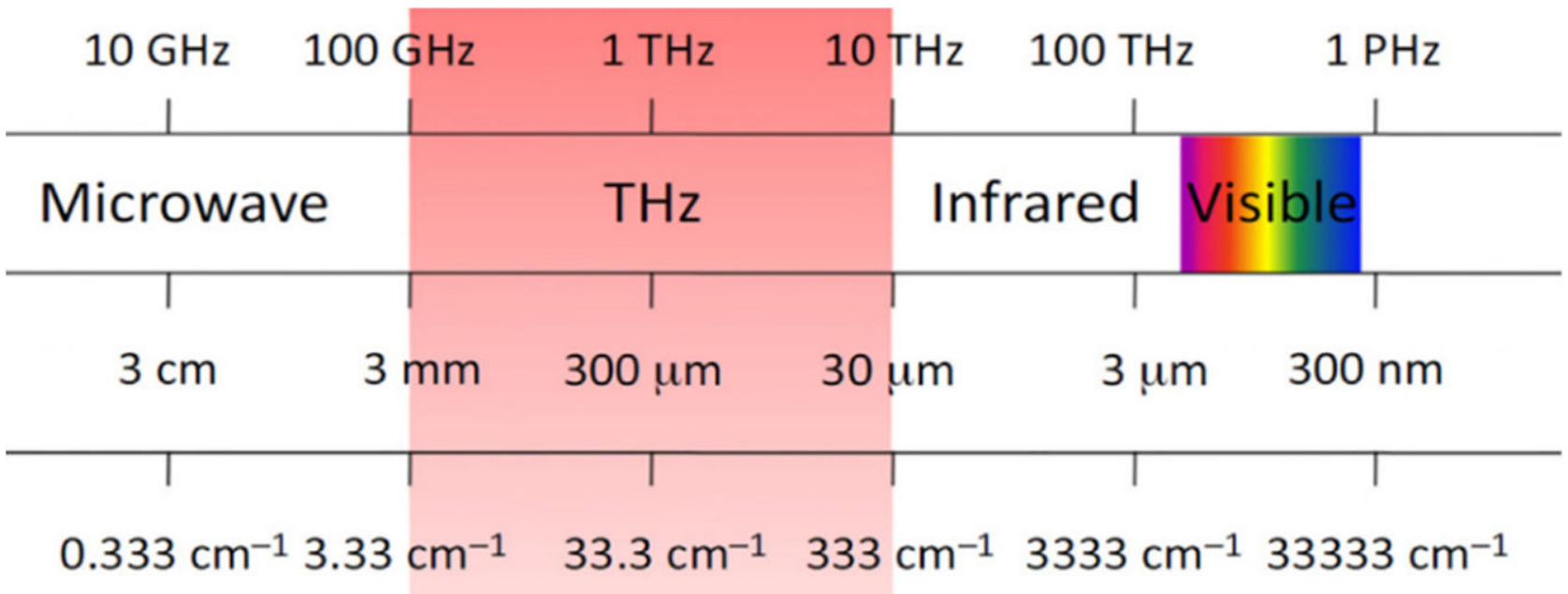

Figure 1: Electromagnetic spectrum of waves [5].

.

A defining advantage of THz waves is their ability to penetrate non-metallic materials, such as skin and clothing, without the ionizing radiation associated with X-rays. This makes them ideal for imaging techniques like THz spectroscopy and tomography. In medicine, THz imaging enables non-invasive diagnosis of skin diseases, including melanoma and other cancers, by detecting changes in tissue structure. It can also monitor wound healing, providing depth and stage information throughout the recovery process [6–8]. In microelectronics, the ultra-high-speed data transmission capabilities of THz waves are promising for next-generation wireless networks, offering massive bandwidth and faster transfer rates. They also enable advanced chip inspection methods, such as THz microscopy, for non-destructive evaluation of integrated circuits, defect detection, and quality assurance [9–11]. Agriculture benefits from THz technology through quality and moisture assessment of crops. By analyzing THz interactions with agricultural products, parameters such as ripeness, freshness, and internal composition of fruits, vegetables, and grains can be evaluated. THz imaging can also detect pests, diseases, and contaminants, while remote sensing applications provide valuable soil property and plant growth monitoring for precision agriculture [12, 13]. In forensic science, THz waves can penetrate materials to reveal concealed objects, identify counterfeit documents, and analyze the chemical composition of drugs and explosives [14, 15]. THz imaging supports non-contact evidence examination, package scanning, and crime scene analysis. Security applications include THz scanners at airports and other high-risk locations for detecting hidden weapons, explosives, and contraband, without the health hazards of ionizing radiation. THz imaging also contributes to cultural heritage preservation [16, 17]. It enables the analysis of fragile artifacts and historical documents without damage, revealing hidden layers, text, and artworks in manuscripts and paintings. This non-invasive method aids restoration efforts while ensuring long-term preservation [18, 19]. Ongoing research continues to expand the capabilities of THz technology, with scientists and engineers developing new techniques, materials, and devices to harness its full potential across a broad spectrum of applications. Recent comprehensive reviews, such as Sun et al. [20], have surveyed THz generation from laser–plasma interactions, emphasizing mechanisms like ponderomotive-driven currents and transition radiation in gaseous and solid targets, with reported high-energy outputs up to 10μJ per pulse and efficiencies approaching $10^{-4}$ in optimized filamentation setups. While sharing some foundational aspects, the present review builds on this by focusing on multi-color (two- and three-color) laser schemes across gas, liquid, and clustered plasmas, providing quantitative models and efficiency comparisons, addressing gaps in clustered media dynamics and relativistic/nonrelativistic regime transitions.

This perspective review focuses mostly on the various key aspects of THz mechanisms in laser-plasma interactions. The generation of THz waves via laser-induced plasma across gas, liquid, and clusters is investigated. Given the complexity of laser-plasma interactions, our focus is narrowed to delving into the THz generation across diverse targets. This discussion encompasses the mechanisms and characteristics of the emitted THz waves, with minimal involvement of ionization dynamics. This paper is organized as follows: The various sources for THz wave generation are investigated in section 2, and their limitations and advantages will be discussed briefly. In Section 3, we will delve into THz wave generation from laser-induced plasma in gas, specifically investigating THz generation from gas plasma induced by two/three-color pulses in both relativistic and nonrelativistic regions. Subsequently, in the following sections, we will explore THz generation from laser-induced liquid plasma and THz generation from laser-induced clustered plasma. These discussions will encompass the THz generation associated with induced plasma in liquid and clustered media, alongside their respective physical mechanisms and endeavors to enhance the performance of the generated THz waves. Finally, we will consolidate these THz generation scenarios and deliberate on the future development and challenges of plasma-driven THz sources in the "Summary and Outlook" section.

## 2 Sources

Generally, there are two distinct types of THz sources, continuous-wave (CW) and pulsed sources. Pulsed THz sources, such as photoconductive antennae, are well-suited for applications that require high peak power, such as nonlinear spectroscopy, time-resolved measurements, and nonlinear imaging. They offer exceptional spectral resolution and support multiple spectroscopic techniques, including THz-TDS. On the other hand, continuous-wave THz sources emit a continuous stream of THz radiation typically operate in the frequency domain, and are based on optical or electronic techniques. The CW THz sources are suitable for applications that require continuous and stable THz output, such as communication systems, THz imaging, and spectroscopy. They provide continuous frequency coverage and are often more compact and easier to operate compared to pulsed THz sources. From the aspect of other divisions and based on the technologies used by adjacent regions in the electromagnetic spectrum, THz radiation can be generated by frequency up-conversion (based on electronic sources) or frequency down-conversion (based on photonic sources) [21]. However, few of them are capable of producing the high peak intensities which is needed for most experiments in nonlinear optics regions. These sources are usually photoconductive antennas or optical rectification in a nonlinear medium that can generate very short pulses of THz radiation, but typically with very low pulse energy- roundly 10 fJ per THz pulse. Other THz sources generate higher average power— vacuum electronic sources such as gyrotron can even reach into the megawatt range—but even in these cases, because the pulse duration is long the achievable peak intensity is partly low, a lot of heat dumped into a medium and the realm of nonlinear optics remains largely out of reach [21, 22].

### 2.1 Photoconductive Antenna

Among the various THz generation techniques, the photoconductive antenna (PCA)—also referred to as a photoconductive switch (PCS)—is one of the most widely used devices. Based on femtosecond laser excitation (Fig. 2), PCAs offer key advantages, including room-temperature operation, broadband emission, and compact design. First introduced for THz generation by D. H. Auston [23], this approach has revolutionized THz generation and detection. A typical PCA consists of a semiconductor substrate with an anode–cathode electrode pair separated by a small gap, across which a bias voltage is applied. When a femtosecond laser with photon energy exceeding the semiconductor bandgap illuminates the gap region, photo-generated electron–hole pairs are created and accelerated by the bias field, producing a transient photocurrent. The electrodes act as antennas, radiating in the THz frequency range, while the carrier density subsequently decreases through recombination. The duration of the transient current—and thus the THz pulse—is determined by the pump laser pulse width and the carrier lifetime. If the photoconductor has a short carrier lifetime, recombination begins immediately after optical excitation, resulting in a short photocurrent pulse and broad THz bandwidth. Conversely, in materials with long carrier lifetimes, photocurrent persists beyond the optical pulse duration, broadening the emitted pulse and reducing bandwidth. To achieve sub-picosecond carrier lifetimes and maintain broad bandwidth, low-temperature-grown gallium arsenide (LT-GaAs) is most commonly used [24–26].

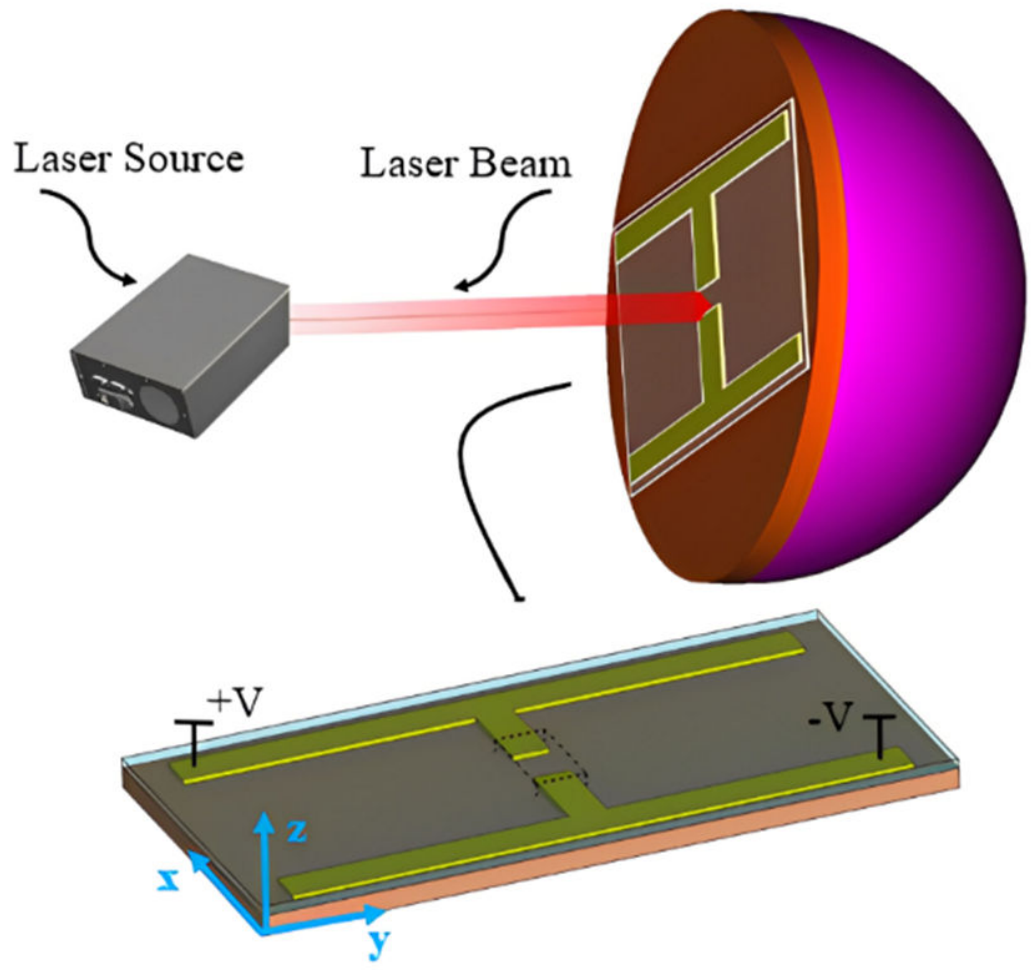

Figure 2: Generation of THz waves by the broadband method on Photoconductive Antenna [30].

In PCAs crucial parameters such as design geometry, doping, growth technique, and choice of semiconductor are parameters to be optimized for maximum THz generation. The overall lateral dimension L of the device is typically in the range of a few millimeters to around a centimeter. The antenna dipole length D is typically on the order of 100 μm, while the gap dimension G can range from a few micrometers to almost D. Nowadays, PCAs can function with nJ level pump laser energy and emit in the 0.3 to 6 THz frequency range. They are used for their compact design and mature technology. PCAs feature a typical energy conversion efficiency of $\sim 10^{-4}$ and focused electric fields of the order of tens of kV/cm. However, limiting behavior arises with a further increase of laser pulse energy as it leads to the breakdown of constituent materials. Increasing high voltage can improve THz flux but degrades the semiconductor base over time [27 - 29].

## 2.2 Backward wave oscillator

A backward-wave oscillator (BWO) is a vacuum electronic device—typically employing a helical or backward-wave transmission line—that can generate high-frequency electromagnetic waves, including those in the THz range (Fig. 3) [31, 32]. The device requires an electron source, most commonly a heated cathode producing thermionic emission [33, 34]. The emitted electrons are accelerated by a high voltage between the cathode and anode and guided through a slow-wave structure (SWS). As the electron beam traverses the SWS, it interacts with the traveling wave, leading to velocity modulation and periodic bunching of electrons [35–37]. This bunching process generates THz radiation via the Smith–Purcell effect, which occurs when a charged particle beam passes close to a periodic structure, such as a grating. In a BWO, the periodic variations of the SWS induce alternating acceleration and deceleration of electrons due to its electromagnetic fields, resulting in coherent radiation emission. The output frequency depends on the SWS periodicity and the velocity modulation of the electron beam, with higher harmonics arising from integer multiples of the fundamental frequency. By tailoring the SWS geometry and optimizing beam parameters, the Smith–Purcell effect can be exploited to produce coherent THz waves. In a BWO, the slow-wave structure supports backward-wave propagation, causing the generated radiation to travel opposite to the electron beam. The THz output is extracted via an output coupler, which transmits the radiation while reflecting residual energy back into the SWS to maximize energy transfer efficiency.

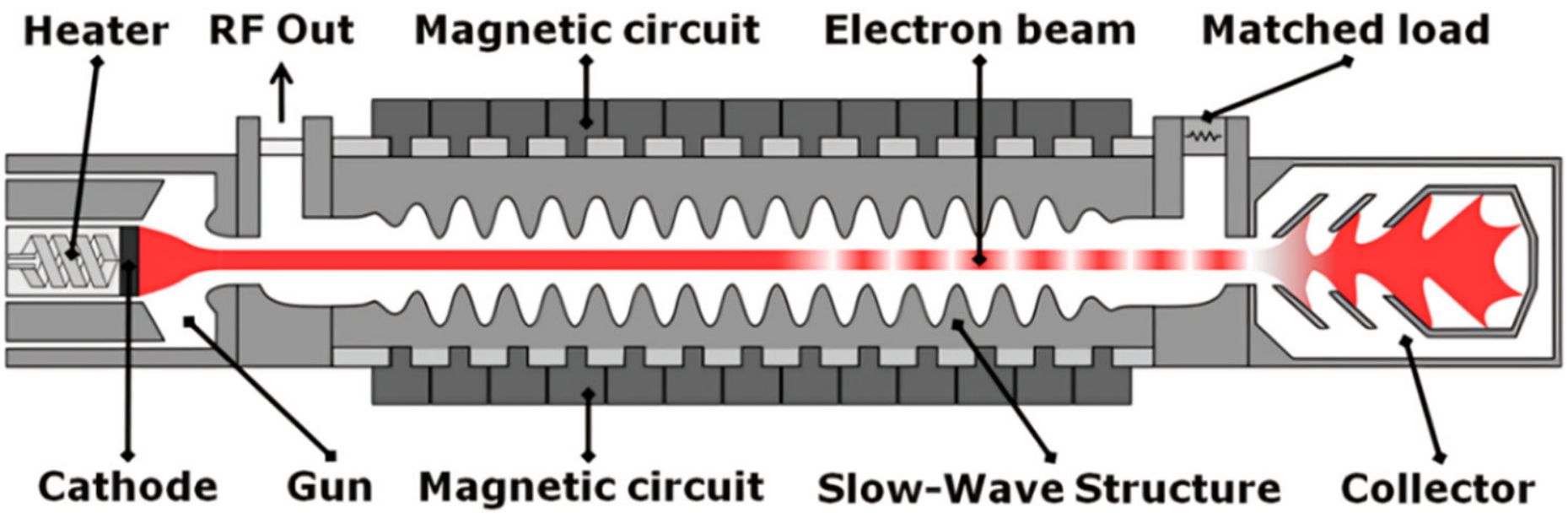

Figure 3: Simplified sketch of a vacuum electron device (left), and a sample dispersion curve of a backward wave oscillator (right) [38].

Despite the BWOs have been successfully implemented for THz radiation and have found applications in various fields, there are also some challenges [39, 40]. The BWOs operate in a high vacuum environment to prevent electron scattering and loss of beam quality. Maintaining a high vacuum can be challenging and requires proper sealing and vacuum pumping techniques. In addition, factors such as beam current, energy spread, and transverse beam size need to be carefully controlled to achieve efficient and stable THz wave generation. These cause ongoing research and advancements continue to address these limitations, leading to improved designs and performance of BWOs for THz wave generation.

### 2.3 Difference frequency generation

In addition to the aforementioned methods, difference frequency generation (DFG) offers a powerful approach for THz wave generation using nonlinear crystals. DFG is a second-order nonlinear optical process in which two input laser beams of different frequencies are mixed in a nonlinear medium to produce radiation at the difference frequency, which can fall within the THz range [41, 42]. Typically, the pump lasers operate in the infrared (IR) or near-infrared (NIR) region, where stable, high-power laser sources are readily available [43]. The nonlinear crystal must possess a non-centrosymmetric structure to exhibit the required χ^(2) nonlinearity [44, 45]. Efficient DFG requires phase matching between the interacting waves to ensure coherent propagation and maximize conversion efficiency. This can be achieved using either birefringent phase matching or quasi-phase matching (QPM) [46, 47]. In birefringent phase matching, anisotropic crystals such as calcite ($CaCO_3$), quartz ($SiO_2$), or rutile ($TiO_2$) are oriented so that the refractive indices for the interacting waves match at a specific phase-matching angle [48, 49]. This requires precise control of the crystal orientation and beam polarization. In QPM, the nonlinear coefficient of the crystal is periodically modulated via periodic poling, compensating for phase mismatch and allowing crystals without natural birefringent phase matching to be used. QPM provides greater material flexibility and supports a broader spectral and temperature range compared to birefringent phase matching.

A variety of organic and inorganic crystals are employed for DFG-based THz generation [22]. Organic crystals often exhibit large second-order nonlinear coefficients and broad transparency in the THz range, enabling high conversion efficiency with minimal absorption losses [41, 50]. However, they generally have lower laser damage thresholds and reduced thermal conductivity. In contrast, inorganic crystals typically offer higher damage thresholds, allowing the use of more powerful pump lasers for increased THz output. Their superior thermal conductivity aids in heat dissipation, making them well-suited for high-power and long-duration operation. The choice between organic and inorganic crystals thus depends on application-specific requirements, including desired THz frequency range, conversion efficiency, power handling, and thermal stability. Here are some examples of commonly used nonlinear crystals for DFG in the THz range, briefly. Lithium Niobate ($LiNbO_3$) is a widely used nonlinear crystal for THz DFG and has a large second-order nonlinear coefficient and good transparency in the THz frequency range. Lithium niobate crystals can be tailored to achieve quasi-phase-matching (QPM) through periodically poled structures, enhancing the efficiency of THz generation. They are suitable for both pulsed and continuous-wave (CW) THz sources [51, 52]. Periodically Poled Lithium Niobate (PPLN) is a variant of lithium niobate where the ferroelectric domains are periodically inverted using electric field poling techniques. This periodic poling enables quasi-phase matching, enhancing the efficiency of THz DFG. PPLN crystals offer versatile and efficient THz generation options and can be tailored for specific THz frequencies [46]. Gallium Selenide (GaSe) is a nonlinear crystal with a large second-order nonlinear coefficient and good THz transparency. It is often used for DFG-based THz sources. GaSe crystals can be phase-matched for specific THz frequencies, and they offer good conversion efficiency and power output [53]. Lithium Tetraborate ($Li_2B_4O_7$) or LBO is a promising crystal for THz DFG due to its high damage threshold, wide transparency range, and large nonlinear coefficient. It offers good phase-matching capabilities and has the potential for efficient THz generation over a broad spectral range [54]. Gallium Orthophosphate ($GaPO_4$) is a crystal with a large second-order nonlinear coefficient and high damage threshold. It has good transparency in the THz range and can be engineered to achieve phase matching for efficient THz DFG. $GaPO_4$ has shown promise for THz generation in recent studies [55]. Potassium Titanyl Phosphate ($KTiOPO_4$) or KTP is a well-known nonlinear crystal used in various frequency conversion processes. Researchers are exploring its potential for THz DFG by optimizing its phase-matching conditions and crystal properties. KTP offers good conversion efficiency and has the advantage of being commercially available [56]. Organic-Inorganic Hybrid Crystals are gaining attention for THz DFG. These crystals leverage the advantages of both organic and inorganic materials, such as large nonlinear coefficients, broad transparency, and high damage thresholds. Examples include organic-inorganic hybrid perovskites and metal-organic frameworks (MOFs) [57]. Chalcogenide Glasses are amorphous materials composed of sulfur, selenium, tellurium, and other elements. They exhibit large nonlinear coefficients and good transparency in the THz range. Researchers are exploring the potential for THz DFG due to their unique properties, including high refractive index, low phonon energy, and adjustable composition.

## 2.4 Laser-Plasma Interactions

Plasma is a fascinating medium for generating THz waves that have owner advantages especially. Plasma-based sources can generate THz waves with a wide bandwidth, enabling the generation of THz radiation over a broad range of frequencies. These can not only support high-intensity fields, but also the properties of plasma, such as density, temperature, and electron energy distribution, can be controlled and tuned to optimize THz wave generation. In addition, plasma-based sources can produce THz pulses with ultrafast rise times that offer promising prospects for applications in imaging, spectroscopy, communications, and security screening due to the unique properties of THz radiation and the versatility of plasma as a medium for its generation and manipulation. Plasma parameters critically influence laser–plasma THz generation and can be partially tuned through laser and target properties, though inherent instabilities limit precision. Electron density, ranging from $n_e \approx 10^{16}$–$10^{19}$ cm$^{-3}$, scales with laser intensity ($I \propto n_e$) due to ionization thresholds, where higher intensities ($I \approx 10^{13}$–$10^{16}$ W/cm$^2$) enhance multiphoton or tunneling ionization. Electron temperature, $T_e \approx 1$–$10$ eV, increases with pulse duration ($\tau \approx 10$–$100$ fs) via inverse bremsstrahlung heating, with absorption coefficient $\alpha_IB \approx n_e^2/\sqrt{T_e}$, reflecting collisional energy transfer. The energy distribution often deviates from Maxwellian, exhibiting suprathermal tails described by $f(E) \approx E^{-\alpha}$, where $\alpha \approx 2$–$3$, driven by nonlinear laser–plasma interactions. Shorter laser wavelengths ($\lambda \approx 400$–$800$ nm) promote tunneling ionization, increasing $n_e$, while target design, such as gas jets creating density gradients ($\nabla n_e/n_e \approx 10^{18}$–$10^{20}$ cm$^{-4}$), optimizes plasma conditions. However, control is constrained by instabilities, including Rayleigh-Taylor instabilities with growth rate $\gamma_RT \approx \sqrt{((g\Delta\rho)/\rho)}$, where g is acceleration and $\Delta\rho/\rho$ is the density contrast, and stochastic filamentation causing density fluctuations $\delta n_e/n_e \approx 10$–$20\%$. These limit shot-to-shot stability to ~10%, as demonstrated in laser-wakefield experiments [23-35]. These parameters and their limitations shape the efficiency and output of THz generation mechanisms, detailed in Section 3.

In the technique for generating THz waves through the interaction of laser pulses with plasma, when a laser pulse strikes a target (whether solid, liquid, or gas), the atoms within the target undergo rapid ionization due to the intense laser field, leading to the formation of a plasma environment. This ionization process occurs when the laser field's intensity exceeds the Coulomb field strength between atomic particles, resulting in plasma formation within the focal volume. Following ionization, the liberated electrons are accelerated by the laser field. The initial observation of THz wave production dates back to 1993 when Hamster and colleagues demonstrated the emission of THz waves through the interaction of monochromatic lasers with plasma [58]. By focusing high-intensity laser pulses onto gas atoms, they observed THz wave pulses emitted from the plasma formed within the focus. The production mechanism, elucidated through ponderomotive force in plasma, involves the rapid separation of electric charges within less than a picosecond, with the resulting ponderomotive current emitting electromagnetic waves in the THz frequency range. The emitted THz wave radiation typically possesses energy levels around 0.1nJ and is limited to several THz frequencies. Another instance of THz wave radiation, as reported by Leemans et al., occurs when a charged particle traverses the boundary between two distinct environments [59]. However, the efficiency of generating THz waves in plasma through monochromatic laser pulses, as demonstrated in the aforementioned experiments, remains notably low, typically around $10^{-6}$ for lasers operating at maximum terawatt power levels.

In all laser–plasma THz generation scenarios, femtosecond laser pulses, with durations $\tau \approx 10$–$100$ fs and intensities $I \approx 10^{13}$–$10^{16}$ W/cm², ionizing a target medium through multiphoton ionization or tunneling ionization, governed by the Keldysh parameter $\gamma=\sqrt{(I_p/(2U_p))}$, where $U_p=(e^2E^2)/(4m_e\omega^2)$ is the ponderomotive energy, $I_p$ is the ionization potential, e is the electron charge, $m_e$ is the electron mass, E is the laser electric field amplitude, and $\omega$ is the laser angular frequency. For MPI ($\gamma > 1$), the ionization rate scales as $w \propto I^k$, where k is the photon order needed to overcome $I_p$, while for tunneling ($\gamma < 1$), the $w \approx \exp(-2(2I_p)^{3/2}/(3eE))$. These processes create a plasma with electron density $n_e \approx 10^{16}$–$10^{21}$ cm⁻³, varying across gases, liquids, or clusters. The ponderomotive force, $F_p = -(e^2/(4m_e\omega^2))\nabla E^2$, where $\nabla E^2$ is the gradient of the squared electric field, drives charge separation, exciting plasma oscillations at the plasma frequency $\omega_p = \sqrt{((4\pi n_e e^2)/m_e)}$. These oscillations generate transient currents $J = -en_e v_ə$, with drift velocity $v_ə \approx eE/(m_e \omega)$, which radiate THz fields through dJ/dt, a source term in Maxwell's equations: $\nabla\times B = \mu_0 J+\mu_0\varepsilon_0(\partial E/\partial t)$. The radiated THz power scales as $P \approx (\mu_0/(4\pi))\int(dJ/dt)^2dV$, proportional to $I_{laser}$ in nonlinear regimes due to enhanced current generation. Key mechanisms include photo-current, where field asymmetry (e.g., from multi-color pulses) induces a net $v_ə \neq 0$, yielding $J_{px} = -en_e v_ə$ and $E_{thx} \propto dJ_{px}/dt$; wave-mixing, driven by third-order nonlinear susceptibility $\chi^{(3)}$, producing $E_{thx} \propto \chi^{(3)}E_x E_{2x} E_{3x}$, where $E_x$, $E_{2x}$, and $E_{3x}$ are fields at fundamental, second, and third harmonics; wakefields from electron bunching, generating $E_{xake} \approx 100$ GV/m; and transition-Cherenkov radiation in denser media (e.g., clusters), where fast electrons ($v_e \approx c$, $\beta = v_e/c$) crossing density gradients produce $E_{thx} \approx (\beta v_e)/(1 - \beta n\cos\theta)$, with n as the refractive index and $\theta$ the emission angle [151]. Media variations arise from electron density: gases ($n_e \approx 10^{16}$–$10^{18}$ cm⁻³) enable long interaction lengths (cm-scale) but yield low efficiency ($\eta \approx 10^{-5}$); liquids ($n_e \approx 10^{19}$–$10^{20}$ cm⁻³, absorption $\alpha \approx 10$ cm⁻¹) achieve $\eta \approx 10^{-4}$ but face propagation limits [164]; and clusters, with Mie plasmon resonances at $\omega_{mie} \approx \omega_p/\sqrt{3}$, boost efficiency to $\eta \approx 10^{-3}$ via localized field enhancements. Multi-color schemes, such as two- or three-color pulses, enhance efficiency by amplifying field asymmetry in photo-current and wave-mixing, achieving $\eta \approx 10^{-5}$ in gases, $10^{-4}$ in liquids, and up to $10^{-3}$ in clusters, driven by plasmonic resonances. Table 1 compares these mechanisms across media, providing quantitative metrics for efficiency, power output, and bandwidth, with multi-color enhancements highlighted. Building on these foundational principles, Section 3 delves into the specific mechanisms of THz generation in laser-induced plasmas, examining their implementation and performance variations across gas, liquid, and clustered media. The detailed analysis will highlight how factors such as electron density gradients, multi-color field asymmetries, and resonance effects influence efficiency and output characteristics, providing a comprehensive exploration of the processes introduced here.

Table 1: Comparison of THz generation mechanisms across gas, liquid, and cluster media, with quantitative metrics for efficiency ($\eta$), power output, and bandwidth.

| Mechanism | Shared Physics | Gas | Liquid | Clusters |
|---|---|---|---|---|
| **Photo-Current** | Asymmetric current | $\eta \sim 10^{-5}$, power 10 nJ–1 μJ, bandwidth 0.1–10 THz | $\eta \sim 10^{-5}$, power 0.5 μJ–5 mJ, high absorption | $\eta \sim 10^{-4}$, power 0.1–10 μJ, resonance-enhanced |
| **Wave-Mixing** | ($\chi^{(3)}$) nonlinearity | $\eta \sim 10^{-4}$, power 0.1 μJ–10 μJ, 0.1–50 THz (2x with three-color) | $\eta \sim 10^{-4}$, power 0.1 μJ–5 mJ, 0.1–30 THz | $\eta \sim 10^{-3}$, power 1 μJ–100 μJ, up to 100 THz with Mie |
| **Wakefield** | Electron bunching | $\eta \sim 10^{-4}$, power 1 μJ–1 mJ | $\eta \sim 10^{-4}$, power 0.1–10 μJ, limited by absorption | $\eta \sim 10^{-4}$, power 10 μJ–1 mJ, enhanced |
| **Transition-Cherenkov** | density gradients | $\eta \sim 10^{-5}$, power 50 nJ–5 μJ, weak | $\eta \sim 10^{-5}$, power 0.1–10 μJ, stronger in liquids | $\eta \sim 10^{-4}$, power 0.5–50 μJ, prominent in clusters |

## 3 Mechanisms of THz Generation

Owing to the rapid development of THz sources and detectors, the understanding of the physical mechanism of its generation has grown significantly. Despite some advantages and limitations of sources as mentioned above, they can not endure high-intensity laser pulses and have material damage. To overcome this obstacle, the use of plasma as an ionized medium that can tolerate very high potential gradients has been of interest. In compresence to other sources of THz radiation, plasma has a great potential to generate high-power broadband, intense, coherent, and highly directional THz waves, and it has become the subject of attention of many researchers in recent years.

### 3.1 THz generation from laser-induced gas-plasma

When intense femtosecond pulses travel through optical media, ionization occurs, leading to the formation of plasma if the optical field strength is high enough to separate electrons. These pulses create waves emanating from the plasma the laser induces in the air. THz generation from air plasma can occur through two distinct types of pump optical fields. Optical pulses drive the first type with a single central wavelength, which is often referred to as a one-color field. This one-color field, first harmonic (FW) typically utilizes ultrafast pulses with a central wavelength of 800 nm. The second type of pump optical field strength is powered by optical pulses with two central wavelengths, known as a two-color field (second harmonic (SHW)). In this case, the pulses have central wavelengths of 800 nm and a second harmonic of 400 nm (Fig. 4). This schematic represents a typical setup for THz wave generation using two-color laser filamentation. Firstly, high-intensity femtosecond laser pulse at a fundamental frequency (ω) is focused using a lens. A Beta-Barium Borate (BBO) crystal is used to generate the second harmonic (2ω) of the laser pulse. The superposition of the fundamental ($\omega$) and second harmonic ($2\omega$) produces a two-color field. The combined beam is tightly focused to create an ionized plasma filament. The asymmetric ionization and transient photocurrents in the plasma filament lead to the emission of broadband THz radiation, which propagates in the forward direction. This technique is widely used due to its simplicity and efficiency in generating high-power THz pulses.

A THz time-domain spectroscopy (THz-TDS) experimental setup is depicted in Fig 5. A Ti:Sapphire femtosecond laser is used as the excitation source. Then the laser beam is split into two paths: the pump beam (to generate THz radiation) and the probe beam (used for detection). A delay stage allows for temporal scanning. The pump beam passes through a BBO crystal to generate a two-color field, which then ionizes a gas medium to produce a plasma that emits THz radiation. The emitted THz pulses are collected and focused onto a sample using parabolic mirrors. A silicon (Si) filter is used to block any residual optical light. The transmitted THz pulse passes through another lens and is detected using an electro-optic sampling method, where the probe beam interacts with the THz field at an electrode-based detector. This setup allows for precise measurement of the THz spectrum and its interaction with different materials. However, the specific properties of the generated THz waves can vary depending on the pump field used. In recent research, scientists have been exploring ways to enhance the efficiency of THz radiation generation. One approach involves considering the third harmonic of the laser pulse (THW), which corresponds to a central wavelength of 267 nm (three times the fundamental 800 nm wavelength). By incorporating the third harmonic, researchers aim to improve the overall efficiency and performance of THz generation systems. The radial THz radiation initiated by interactions of laser-plasma can be described through two main mechanisms. The first mechanism is the four-wave mixing model, which involves the non-linear polarization scheme (Kerr effect). The second mechanism is the photocurrent model, in which the scenario operates on tunneling ionization principles, eliciting linear and nonlinear current densities within the plasma. One scheme for THz radiation generation is the wave mixing (WM) mechanism proposed by J. Penano et al [60]. The scheme entails the simultaneous propagation of first and second-harmonic laser pulses within the plasma, showcasing significant promise for generating broadband high-power THz waves. Four-wave mixing (FWM) in plasma is an approach that is particularly effective for relatively high laser intensities in orders of ($10^{15} - 10^{17} W/cm^2$). The THz radiation generated by the interaction of plasma and sufficiently high laser intensities $\geq 10^{18} W/cm^2$ must also include the relativistic modifications. In this approach, THz radiation power is enhanced due to simultaneous guidance with low loss, making it suitable for communication applications. Another scheme, the photocurrent model (PC), also known as tunneling ionization, for the two-color laser's interaction with plasma, particularly at relatively low laser intensities ($\leq 10^{15} W/cm^2$) is proposed by Kim et al [61, 62]. In this scheme, the nonzero drift velocity of ionized electrons triggers the generation of an induced current density, capable of producing THz radiation within the low-frequency range. Theoretical and experimental investigations conducted by Andreeva et al. showed that the photocurrent scheme is most suitable for explaining the low-frequency segment of the THz spectrum. However, it does not fully elucidate the effective factors influencing THz emission. Conversely, the wave-mixing scheme accurately characterizes the high-frequency range of the THz spectrum and provides several advantages such as a simpler experimental setup, higher spectral purity of the generated radiation, and the ability to explain the effect of each factor that influences the generation of THz radiation.

#### *3.1.1 Photo-Current*

The photo-current mechanism in the context of two-color and three-color laser pulses interacting with plasma involves the generation of electric currents induced by the interaction of pulses with plasma. Temporal frequency beating ($\Delta\omega=2\omega-\omega$) between multi-color components induces ponderomotive gradients ($\nabla U_p \approx (e^2/(2m_e\omega^2))\nabla E^2$), creating spatial charge separation ($\delta n_e \approx -(\varepsilon_0/e)\nabla\cdot(\nabla U_p/\omega_p^2)$) that drives asymmetric electron drifts $v_d \approx eE/(m_e \nu_{en})$, with collision frequency $\nu_{en}$). This forms photo-currents $J_{pc} = -e\delta n_e v_d$, serving as the low-frequency THz radiator with $E_{THz} \propto \int J_{pc} \exp(i\, k\cdot r)\, dV$ [65–67]. In the case of two-color laser pulses, where two distinct wavelengths of light interact with the plasma, the nonlinearity of the plasma medium causes the generation of harmonics and other nonlinear optical effects. When these harmonics are absorbed by the plasma, they can create a spatially varying electric field. This varying electric field, in turn, induces a photo-current in the plasma. The generated photo-current is a response to the intensity and frequency characteristics of the incident two-color laser pulses. Additionally, studies have shown that different frequency ratios of the two-color pulses significantly impact the efficiency and polarization characteristics of the generated THz radiation [68, 69]. In the case of three-color laser pulses, the complexity increases due to the interaction of three different wavelengths of light. The resulting combination of frequencies leads to the generation of a spatially varying electric field within the plasma, inducing a photo-current. The photo-current in three-color interactions is influenced by the nonlinear interplay of the three incident laser wavelengths and the resulting complex plasma response. Recent theoretical and numerical studies suggest that by carefully tuning the phase differences between the three color components, one can optimize the amplitude and directionality of the generated THz waves [70]. Understanding the photo-current mechanism in two-color and three-color laser pulse interactions with plasma is crucial for applications such as laser-driven particle acceleration, THz wave generation, and plasma-based light sources. Researchers study and manipulate these interactions to harness the unique properties of laser-induced plasma for diverse scientific and technological purposes. Due to this, the total laser pulse containing the FW, SHW, and THW with the different polarizations can be written as follows [71, 72]:

$$\vec{E}_L(z,t) = \vec{E}_1(z,t) + \vec{E}_2(z,t) + \vec{E}_3(z,t)$$
$$= \frac{E_0}{2}\left[\left(\hat{e}_x + i\alpha_1\hat{e}_y\right)e^{i(k_1 z-\omega_0 t)} + \left(\hat{e}_x + i\alpha_2\hat{e}_y\right)e^{i(k_2 z-2\omega_0 t)} + \left(\hat{e}_x + i\alpha_3\hat{e}_y\right)e^{i(k_3 z-3\omega_0 t)}\right] \quad (1)$$

where $E_0$ is the amplitude of laser fields. Incorporating the third harmonic (THW, λ=267nm) into the driving laser field enhances THz generation efficiency by amplifying field asymmetry, leading to stronger nonlinear currents ($J_{nl} \propto E_\omega E_{2\omega} E_{3\omega}$) and phase-dependent modulation (Δφ optimization yields $\eta \sim 2 - 3\times$ two − color). Experiments report up to twofold amplitude increases and spectral broadening to >50 THz through coherent control [73]. The parameters of $\alpha_1$, $\alpha_2$ and $\alpha_3$ are defined as the ratio of the electric field amplitudes in x-y plane for various harmonics. Through the adjustment of these parameters, the laser fields polarizations can be regulated relative to one another. With three-color fields the molecule is ionized, and the tunneling ionization rate given by the Perelomov–Popov–Terent'ev (PPT) model [74], which forms the foundation for the commonly cited Ammosov–Delone–Krainov (ADK) approximation (see [75] for historical context and derivations):

$$w(t) = 6.6 \times 10^{16} \frac{Z^2}{n_{eff}^2} \left( \frac{32.61 Z^3 E_H}{3 n_{eff}^4 E_L} \right)^{n_{eff}-1.5} \left( -\frac{2 Z^3 E_H}{3 n_{eff}^4 E_L} \right) \tag{2}$$

here $E_H = 5.14 \times 10^9 (V/cm)$ is the electric field in the atomic unit, $Z$ is the electric charge of ionized atoms, $E_L$ is the laser pulse electric field, $n_{eff} = Z/\sqrt{E_{ion}(eV)/13.6}$ is the effective number of ions where $E_{ion}$ is the ionization rate potential. The radiated THz field is determined by evaluating the induced current density of electrons, denoted as $\vec{J}_{THz}$, where can initiate this process by considering the following equation [40]:

$$\frac{\partial \vec{J}_{THz}}{\partial t} + \nu_e \vec{J}_{THz} = \frac{\omega_p^2}{4\pi} \vec{E}_L \tag{3}$$

where $\omega_p$ is the frequency of plasma and $\nu_e$ is the collision frequency. Due to the THz signal generated by a single-color pulse-induced plasma is relatively weak, numerous efforts have been made to amplify the THz amplitude/conversion efficiency by using an external DC magnetic field, adjusting the interaction length, optimizing the pump pulse focusing conditions, considering the wavelengths/harmonics components of the laser pulse, and manipulating other laser pulses optical parameters [76 - 78].

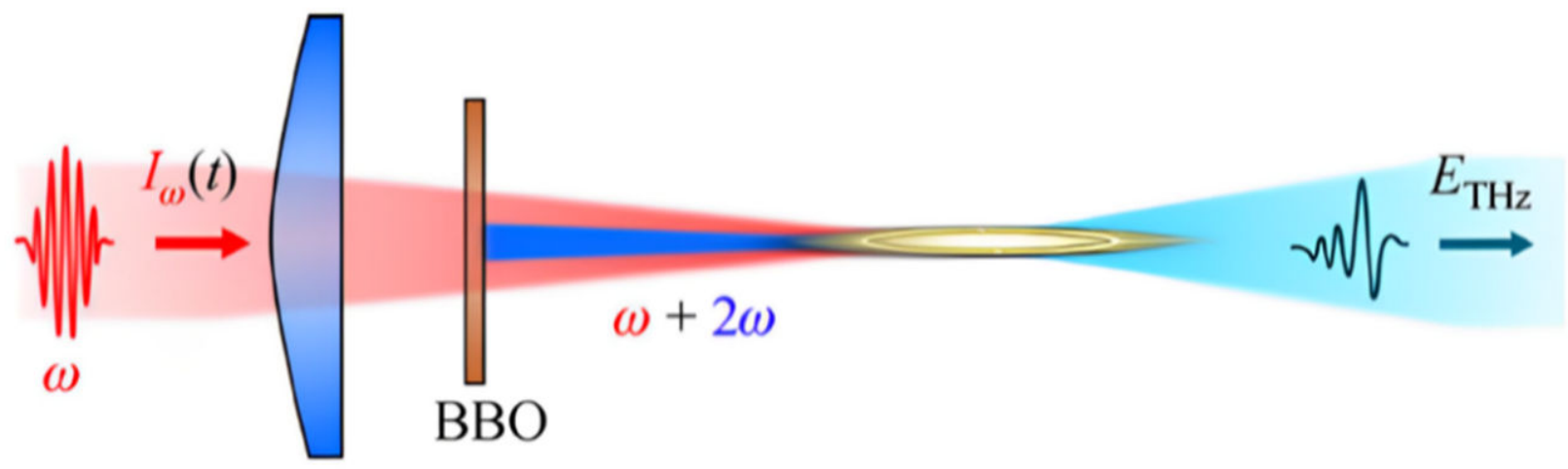

Figure 4: The experimental setup for THz generation from two-color laser-induced air plasma [63].

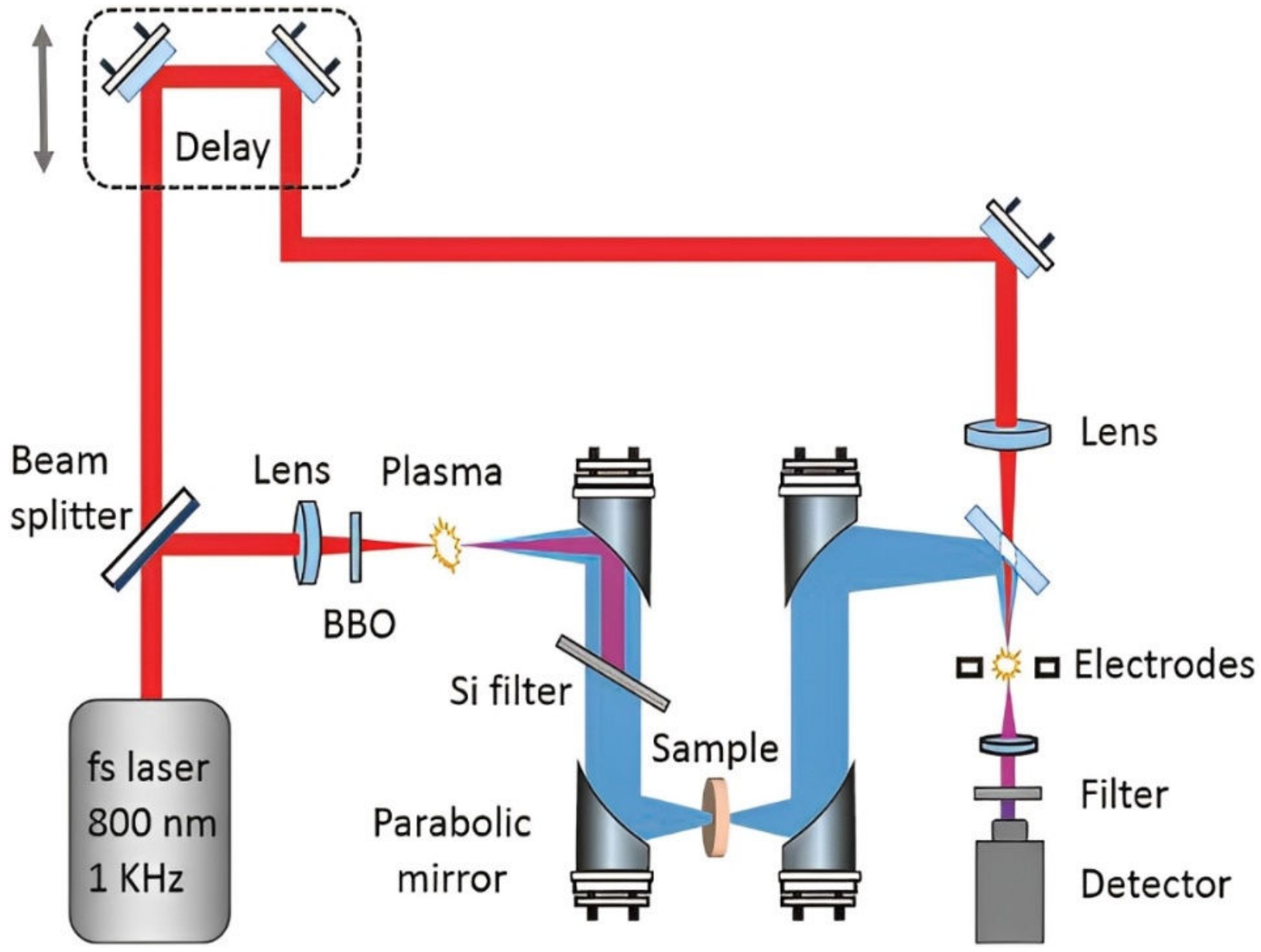

Figure 5: Illustration of the broadband air-plasma THz system [64].

The presence of an external magnetic field can have several effects on the generation and properties of THz radiation through the photo-current mechanism [79 - 81]. An external magnetic field B alters plasma optical properties through the Lorentz force, inducing cyclotron motion at ω_c = eB/$m_e$ (≈$10^{11}$ rad/s for B = 1T) and effects like Faraday rotation or Zeeman splitting. These enhance THz efficiency via resonant absorption and modified dispersion ($n_{eff} = \left(1 + \omega_p^2\right)/(\omega^2 - \omega_c^2)$), increasing output by $\times 2 - 5$ in experiments. However, the effects of it in the photo-current mechanism depend on the specific experimental setup and the materials' characteristics involved. When a material is subjected to a magnetic field, its optical properties are altered [83]. This includes changes in the refractive index, absorption coefficient, and emission characteristics. The presence of a magnetic field influences the recombination dynamics of photo-excited charge carriers, the spin properties, and the spatial distribution of the charge carriers, altering their recombination pathways and lifetimes. As a result, the emission intensity and temporal characteristics of the generated THz radiation be modified. The external magnetic field modifies the optical properties of the material, leading to changes in the absorption and emission of THz radiation. This can alter the efficiency and spectral characteristics of the photo-current mechanism. Furthermore, in a magnetic field, charged particles experience a force, the Lorentz force, perpendicular to their motion [84 - 89]. This force causes the charged particles to move in circular orbits with a frequency called the cyclotron frequency. If the frequency of the incident THz radiation matches the cyclotron frequency of the charge carriers in the material, it leads to cyclotron resonance. Since it facilitates the resonant excitation of charge carriers, cyclotron resonance significantly enhances the efficiency of THz generation through the photo-current mechanism.

The higher harmonic components of the laser pulse can lead to enhanced THz generation efficiency in ordinary and extraordinary modes [90-92]. The intensity of the higher harmonic components is typically lower compared to the fundamental frequency, but their contribution to THz generation can still be significant. By including higher harmonic components in the laser pulse, the interaction with the material can be optimized, resulting in more efficient acceleration or deceleration of charge carriers and enhanced THz radiation emission. In addition, the presence of higher harmonic components in the laser pulse can broaden the spectral content of the generated THz radiation. Each harmonic component contributes to the overall THz emission spectrum, resulting in a broader frequency range. This broadening can be advantageous for applications that require a wide bandwidth of THz radiation. Optimizing the focusing condition of the pump pulse requires careful consideration of the material properties, pump pulse parameters (such as wavelength, duration, and intensity), and experimental setup. Experimental investigations and numerical simulations are commonly employed to determine the optimal focusing conditions for achieving efficient and controlled THz radiation generation in the photo-current mechanism [93- 95]. The focusing condition of the pump pulse plays a crucial role in determining the spatial distribution and intensity of the photo-generated charge carriers. By optimizing the focus of the pump pulse, the concentration of the pump energy can be maximized in the active region of the material, and the spatial distribution and intensity of the photo-generated charge carriers can be modified, leading to a more enhanced THz generation efficiency. Furthermore, by properly focusing the pump pulse, the direction of the generated THz radiation can be controlled. Tighter focusing can result in more directional THz emission, whereas de-focusing or using a larger beam diameter can lead to broader THz emission patterns. This control over the THz emission directionality can be crucial for specific applications that require precise targeting or beam steering.

### *3.1.2 Wave-Mixing*

The experimental conditions for THz generation from gas plasma induced by a one-color pulse are relatively straightforward, as demonstrated by Hamster et. al. In this scenario, a formed plasma is sufficient to facilitate the generation process [96 - 98]. During this experiment, a laser pulse with a pulse width of 120 fs and energy of 50 mJ was focused to generate plasma, subsequently, pulsed radiation of electromagnetic waves at THz frequency was observed, as depicted in Fig. 6. However, the efficiency of this configuration is relatively low, and to enhance efficiency, researchers proposed exploring the interaction of two-color and three-color lasers with plasma. The experimental setup (Fig. 6a) utilizes a Ti:sapphire laser amplifier to generate femtosecond pulses, which pass through an optical parametric amplifier (OPA) and a BBO crystal to produce a three-color field ($\omega + 2\omega$). This mixed field ionizes a gas medium, creating a plasma filament that efficiently emits broadband THz radiation. Optical components such as a temporal delay stage, half-wave plate, and parabolic mirrors optimize the process, while a silicon filter removes unwanted optical signals. THz detection is performed via electro-optic (EO) sampling with a ZnTe crystal. The results (Fig. 6b) compare the normalized THz temporal waveforms generated by two-color and three-color fields, showing that the additional frequency component significantly enhances THz intensity. Similarly, (Fig. 6c) demonstrates that the three-color scheme modifies the waveform structure, indicating improved control over THz pulse characteristics, which is beneficial for applications in spectroscopy and imaging.

In the case of two-color laser pulses, the interaction involves two distinct wavelengths of light. Nonlinear processes, such as harmonic generation and sum or difference frequency generation, occur due to the combined effect of these wavelengths interacting with the plasma [99]. These interactions lead to the creation of new frequencies that are multiples or combinations of the original wavelengths [100, 101]. When three-color laser pulses interact with plasma, the nonlinear effects become even more intricate. The three different wavelengths of light interact in a nonlinear manner, leading to a wider range of possible interactions. Four-wave mixing, where three incident waves combine to generate a fourth frequency, is a significant process in three-color interactions. Through this process, various combinations of the three input wavelengths result in the generation of additional frequencies, enabling the production of a rich spectrum of electromagnetic waves [102]. The wave-mixing mechanism in both two-color and three-color laser pulse interactions with plasma relies on the nonlinear properties of the plasma medium. These interactions are essential in fields such as spectroscopy, laser-induced plasma physics, and the generation of new frequencies for applications in areas like THz technology and laser-driven particle acceleration. Understanding and controlling these mechanisms is crucial for harnessing the potential of laser-plasma interactions in diverse scientific and technological applications. the specific details of the wave-mixing mechanism for THz radiation generation using three-color laser pulses and plasma can depend on the experimental setup and the characteristics of the laser pulses and plasma medium used. Advanced theoretical and experimental studies are typically conducted to optimize these parameters and understand the underlying physics of the process. Plasma absorption can affect the propagation of the laser pulses through the plasma and, consequently, impact the efficiency of the wave-mixing process. Higher plasma densities and temperatures can result in increased absorption of the laser pulses. This absorption can lead to energy loss and reduced THz radiation output. The density of the plasma affects the efficiency of the wave-mixing process and the strength of the THz radiation generated. Higher plasma densities generally lead to stronger nonlinear interactions between the laser pulses and the plasma, resulting in enhanced THz radiation. However, there is an optimal density range where the interaction is most efficient. If the plasma density becomes too high, it can lead to excessive absorption or scattering of the laser pulses, reducing the THz radiation output. Therefore, controlling the plasma and laser parameters, including the intensities and wavelengths of the laser pulses, density and temperature of plasma, is crucial to balance the absorption and efficient conversion of optical frequencies to THz frequencies.

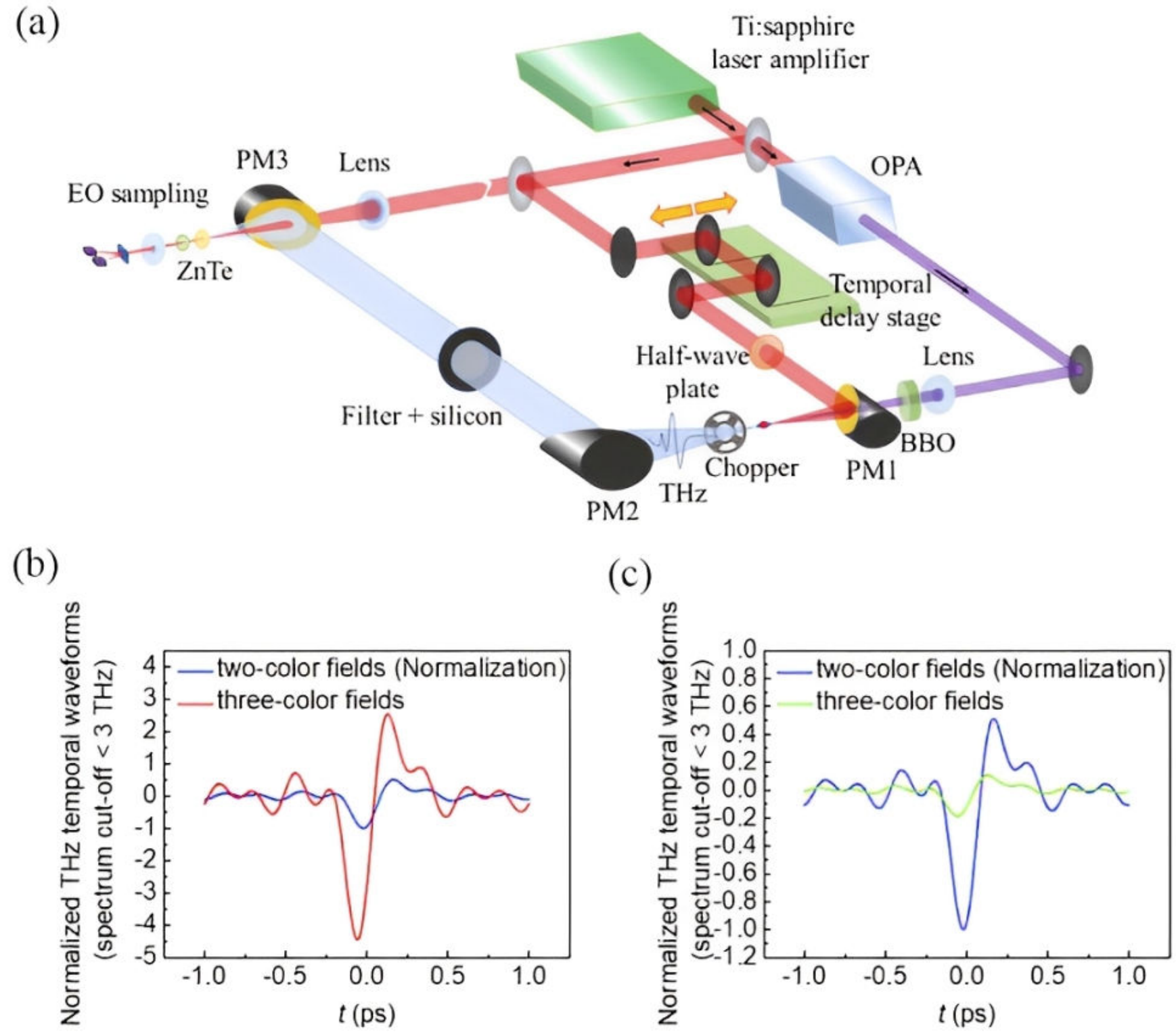

Figure 6: (a) Diagram illustrating the experimental setup, (b) THz temporal waveforms under ideal conditions, where the two plasmas perfectly overlap, (c) THz temporal waveforms under real conditions, where the two plasmas are not completely aligned [103].

In the field of optical ionization and THz radiation in the interaction of two-color and three-color of pulses with plasma, several studies have been conducted to understand the effects of different laser pulse configurations. Andreeva and colleagues investigated the generation of THz radiation through femtosecond laser-induced plasma [65]. They developed a theoretical model that considered the contribution of free electron current and nonlinear polarization of molecules. Their results showed that the photocurrent mechanism plays a significant role in generating the low-frequency part of the THz spectrum, while the higher-frequency part can be explained by Kerr's nonlinear response of molecules. Fedorov et al. compared four-wave mixing and photocurrent models for THz generation by two-color femtosecond filaments [104]. They studied the specific polarization conditions under which each mechanism is responsible for THz generation and found that the predictions of the two models are qualitatively similar. However, they noted that the photocurrent model lacks precise mathematical details and may not accurately account for all factors influencing THz radiation. Song Li and colleagues focused on laser-plasma acceleration using two-color relativistic femtosecond laser pulses [105]. They conducted experiments and numerical simulations to study the second harmonic generation process and its impact on sustaining the acceleration structure in the plasma over longer distances. Manendra et al. focused on bright THz wave generation through frequency mixing of laser pulses in inhomogeneous cold plasma [106]. They introduced a theoretical framework for generating radially polarized THz waves and showed that by adjusting the profile parameters of the laser pulse, and one can tailor both the efficiency and the distribution of the field. S. Liu et al. explored the manipulation of boosted THz emission from plasma using a femtosecond three-color laser fields [107].

They observed that the amplitude of the THz wave emitted by the three-color wave is roughly doubled when compared to the two-color scheme. Z. Zhou et al. conducted quantum calculations to analyze THz generation using multi-color intense laser fields with different frequency ratios [108]. According to their model, the enhancement in THz radiation in multi-color schemes is predicted to happen when the multi-photon mixing condition is met. H. Wang and workers investigated the generation and evolution of various THz one beams from plasma filaments [109]. They introduced a theoretical and numerical model centered around a shaped two-color laser field, elucidating that the plasma channels induced by this field stem from four-wave mixing processes. The progression of these channels depends upon the interplay of the pump time delay and the distinct characteristics of periodic and helical plasma filaments. W. Sheng and workers investigated the spectral control of THz fields in inhomogeneous cold plasma channels by customizing two-color laser beams [110]. Their experimental findings demonstrated that the THz wave emitted in the far-field region displays coherent emission stemming from different regions of the inhomogeneous plasma densities, enabling the manipulation of both the energy and intensity distribution of the THz wave fields.

In the framework of the wave-mixing model and interaction of three-color laser pulses with plasma, based on Eq. 1, the laser fields steer plasma electrons to drift in the direction of laser fields and the electron's motion induces a nonlinear current density along the laser electric field. By utilizing this induced nonlinear current density, we can assess the THz electric field utilizing the wave equation (Fig. 7) and explore the radiated power and angular distribution of the THz wave in this process (Fig. 8). Consequently, the total energy per solid angle per frequency can be derived to accomplish this objective as follows [72]:

$$\frac{\partial^2 W}{\partial\Omega\partial\omega} = \frac{c^2R^2}{2\pi}\hat{E}\hat{E}^*[(j_{2x}^2 + j_{3x}^2)(1 - Sin^2\theta Cos^2\varphi) + \left(j_{2y}^2 + j_{3y}^2\right)(1 - Sin^2\theta Sin^2\varphi) + (j_{2z}^2 + j_{3z}^2)Cos^2\theta] \quad (4)$$

And so on, for interactions of two-color laser pulses with plasma in the nonlinear induced electron current density of the THz wave at relativistic and nonrelativistic regimes, the time integration yields the total energy per frequency per solid angle evaluated as a function of the plasma and laser pulse characteristics as follows [111 - 113]:

$$\frac{\partial^2 W_{THz}}{\partial\omega\partial\Omega} = \frac{cR^2}{2\pi}p(\nu_e, \omega_c, \alpha)[1 - sin^2\theta sin^2\varphi]|\tilde{E}_{THz}|^2 \quad (5)$$

These equations demonstrate that the interplay between these parameters plays a crucial role in determining the overall behavior and properties of the generated THz wave, and how the THz wave is distributed angularly and its waveform is heavily influenced by the specific characteristics of the plasma and the laser pulse such as cyclotron frequency, length of plasma, intensity of laser pulse, collisions frequency and so on [114, 115]. The presence of a magnetic field affects the motion of charged particles in the plasma, causing them to gyrate around the field lines at the cyclotron frequency. This gyration can lead to the emission of THz radiation through a process of cyclotron resonance. By adjusting the strength of the magnetic field, one can tune the cyclotron frequency and control the THz emission. The length of the plasma medium used can excite plasma waves, such as electron plasma waves or Langmuir waves, which are responsible for THz generation. The length of the plasma determines the interaction time between the laser pulse and the plasma, affecting the efficiency and characteristics of THz generation. Longer plasma lengths allow for a more extended interaction time, potentially leading to enhanced THz radiation. The intensity of the laser pulse directly affects the acceleration of plasma electrons and the resulting THz emission. Higher laser intensities can cause stronger ponderomotive forces, which push the electrons and accelerate them collectively. This acceleration leads to the generation of THz waves with higher field strengths and broader spectral bandwidths. Additionally, collisions between charged particles in the plasma can significantly impact the dynamics and energy transfer processes, affecting the THz generation and reducing the overall efficiency of THz generation.

### *3.1.3 Wakefields*

Laser-wakefield accelerators (LWFAs) have emerged as novel sources of THz radiation by utilizing the interaction between lasers and plasma to accelerate electrons [116]. The concept of LWFAs was first proposed by Tajima and Dawson in the 1970s, introducing the use of the ponderomotive force as an accelerating mechanism [117].

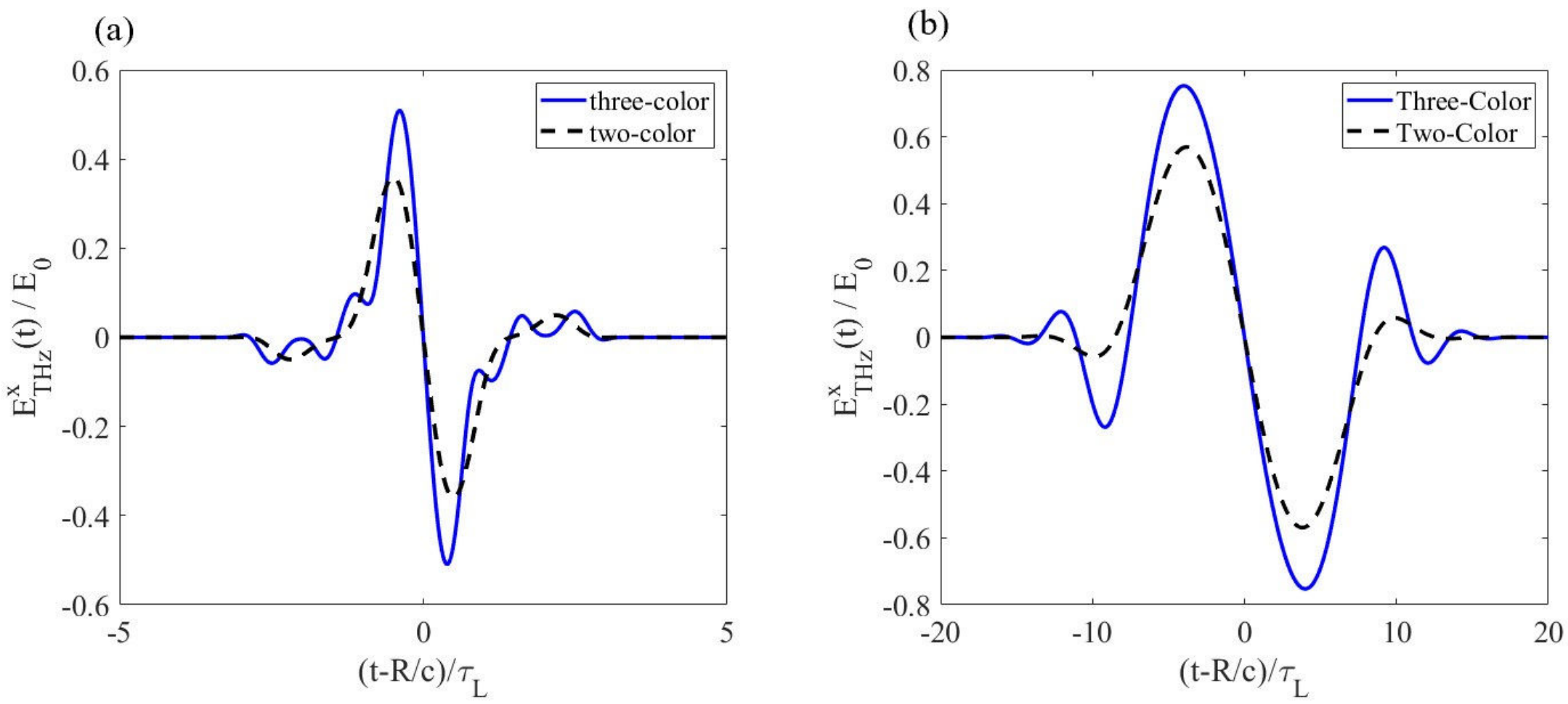

Figure 7: The normalized THz electric field as the normalized retarded time across the pulse duration, showcasing (a) the photo-current mechanism and (b) the wave-mixing mechanism [72].

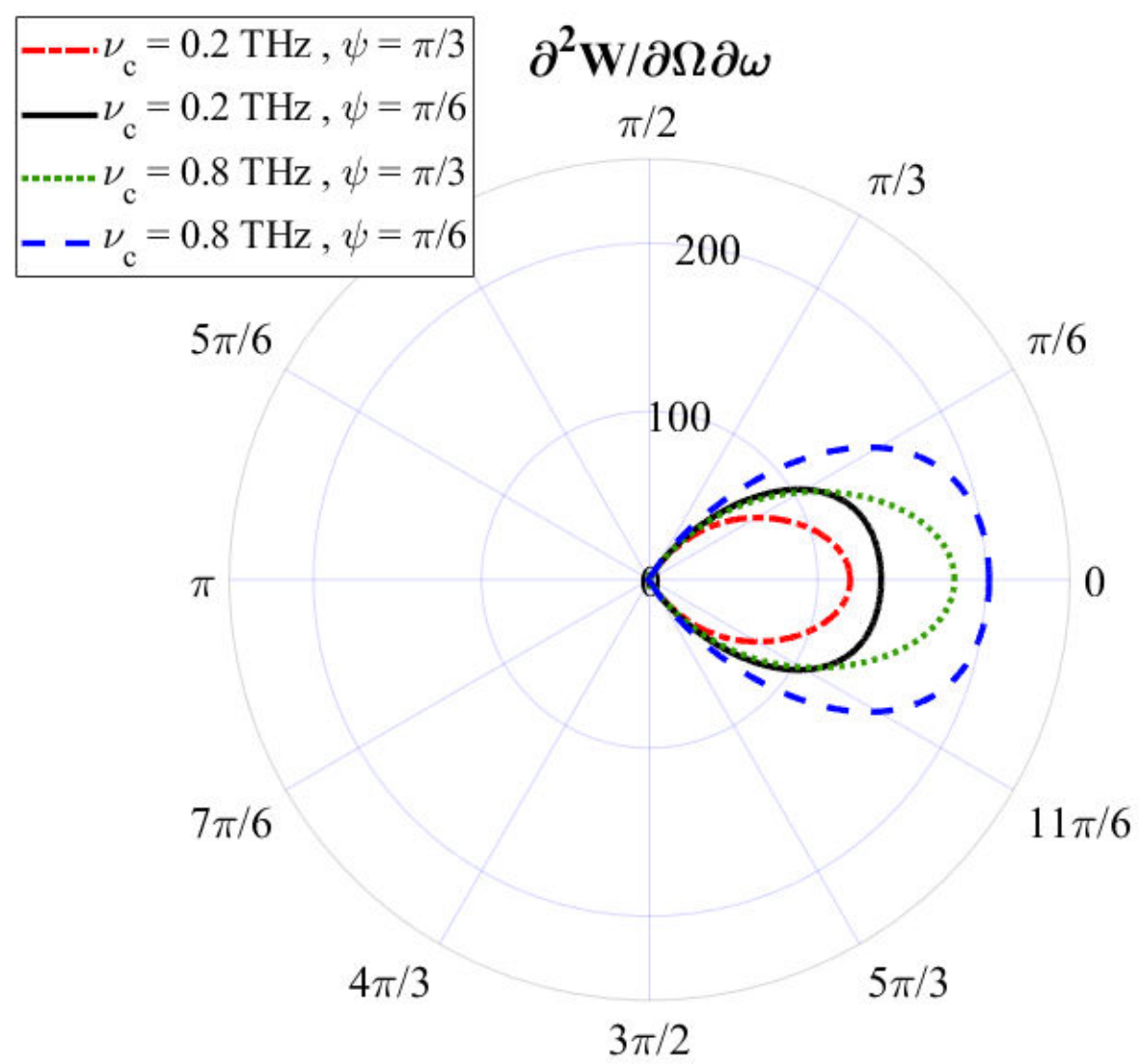

Figure 8: The influence of the oblique angle of the external magnetic field and cyclotron frequency on the angular dispersion of THz wave generated in the context of the wave-mixing mechanism [72].

This pioneering idea has paved the way for significant advancements in accelerator technology. A brief yet powerful laser pulse, as it travels through plasma and leads to charge separation, can generate electric fields that significantly surpass what traditional radio-frequency accelerators can achieve [118-120]. This phenomenon has driven the emergence of laser-plasma wakefield accelerators (LWFAs). The LWFAs are compact devices that exploit the interaction between intense laser pulses and plasma to produce ultra-short electron bunches. When a high-intensity laser pulse traverses low-density plasma, it has the potential to produce electron bunches lasting only femtoseconds. These self injected bunches carry picocoulomb charges and can undergo significant acceleration along the central axis, with their energies depending on both laser intensity and plasma density. This process results in the creation of an accelerating structure formed in the wake of the laser-pulse-plasma interaction, facilitating such electron acceleration. This technology allows for the development of compact tabletop accelerators with numerous applications. One of the key advantages of LWFAs lies in the ability of plasmas to sustain electric fields of up to 100GV/m, which far exceeds the capabilities of traditional accelerator technologies. This high electric field strength has fueled the development of LWFAs as a promising alternative for achieving substantial acceleration gradients. Concurrently, there has been a surge in research efforts dedicated to developing THz sources using laser- and accelerator-based methods [121 - 124]. A persistent challenge in this field pertains to the achievement of narrow-band THz sources capable of delivering pulse energies exceeding the 1μJ threshold. These approaches frequently utilize ultrafast laser pulse systems or particle accelerators to generate THz radiation, harnessing diverse mechanisms like photoconductive antennas and nonlinear optical processes. However, a common observation is that many of these techniques produce isolated cycles of THz waves characterized by heightened amplitudes. The development of LWFAs as THz radiation sources has opened up new avenues for research and technological innovation. Scientists and engineers are actively exploring the optimization of LWFAs to enhance the efficiency, stability, and controllability of THz radiation generation. This includes advancements in laser technology, plasma density control, and electron beam manipulation techniques.

In this respect, coherent transition radiation (CTR) is proposed mechanism to explain the radiation characteristics of THz waves generated by LWFAs [125 - 127]. In this, an electron beam is made to cross an interface between two different media, typically a plasma-vacuum boundary. As the electron beam interacts with the interface, it emits coherent radiation in the THz frequency range that was reported by Leeman et al and other works [59, 128]. They conducted the first experimental demonstration of THz generation from femtosecond electron bunches at a plasma-vacuum boundary. In their experiment, electron bunches with a charge of 1.5 nC were used, and they observed the emission of THz radiation with an energy of 3-5 nJ per pulse. The femtosecond electron bunches used in LWFAs are extremely short, typically on the order of a few femtoseconds. This short pulse duration allows for the generation of ultrashort pulses of THz radiation, which can be useful for various applications requiring high time resolution. In this model, the energy radiated by a bunch of electrons traveling normally to the plasma-vacuum boundary is given by:

$$\frac{\partial^2 W_e}{\partial\omega\partial\Omega} = \frac{r_e}{\pi^2} m_e c N_e (N_e - 1) sin^2\theta \left| \int dug(u) F(\omega,\theta,u) \frac{u(1+u^2)^{1/2}}{1+u^2 sin^2\theta} D(\omega,\rho,\theta,u) \right|^2 \qquad (6)$$

where Boltzmann's distribution can describe the momentum distribution of electron beam. Based on this formula, the properties of the laser pulse used to drive the LWFA play a crucial role in THz generation, and the optimization of LWFA parameters can significantly enhance the efficiency and control of THz generation [124 - 127]. Shorter laser pulses enable the production of ultrashort THz pulses, while higher laser energies provide more energy to the accelerated electron bunch, resulting in increased THz radiation intensity. Proper focusing of the laser beam is important to achieve a well-controlled interaction between the laser and the plasma, leading to efficient electron acceleration and subsequent THz emission. The quality of the accelerated electron beam has a direct impact on the characteristics of the emitted THz radiation. Higher beam charges lead to increased THz radiation intensity while minimizing the energy spread and bunch duration allowing for the generation of narrower and more coherent THz pulses. Controlling the divergence and emittance of the electron beam helps to achieve focused and well-defined THz radiation. Additionally, the design of the plasma-vacuum interface, where the electron beam crosses, is a critical factor in THz generation. This includes engineering the shape, material, and roughness of the interface to control the radiation emission properties. By tailoring the interface design, researchers can enhance the efficiency of THz generation and achieve better control over the emitted THz pulses.

In order to achieve the generation of more intense THz waves using LWFAs, it is crucial to overcome several technical challenges. These challenges stem from the unique characteristics of LWFAs and the requirements for efficient THz radiation generation [131 - 133]. One of the challenges is associated with the low electrical charge of the electron bunches generated in LWFAs. The interaction between the laser and plasma produces electron beams with relatively low charges compared to conventional accelerators. This poses a challenge because the energy of the generated THz radiation is proportional to the square of the electron bunch charge. Therefore, increasing the bunch charge becomes essential for enhancing the maximum intensity of the THz waves produced by LWFAs. Another technical challenge lies in achieving the required sub-micron precision in the alignment of the electron beam. The alignment precision is crucial to ensure optimal interaction between the laser and plasma, maximizing the acceleration efficiency and the subsequent generation of THz waves. Precise control over the position and trajectory of the electron bunch is necessary to achieve the desired THz radiation characteristics. Additionally, the short wavelength of operation in the THz range introduces sub-femtosecond timing requirements [134, 135]. The extremely short timescales associated with THz radiation necessitate precise synchronization between the laser pulses and the electron bunch to ensure coherent emission. Maintaining this synchronization with high precision is a technical challenge that needs to be addressed in the design of THz sources based on LWFAs. Overcoming these technical challenges requires careful consideration in the design and implementation of THz sources based on LWFAs. Researchers are actively exploring various strategies to tackle these issues. This includes advancements in laser technology to produce higher charge electron bunches, the development of precise beam alignment techniques, and the use of sophisticated timing control mechanisms to achieve the required synchronization. By addressing these technical challenges, researchers aim to enhance the intensity and quality of THz waves generated by LWFAs. Overcoming these limitations will enable the realization of more powerful and efficient THz sources, unlocking new opportunities in fields such as materials science, spectroscopy, imaging, and communication.

#### *3.1.4 Transition-Cherenkov*

Contrary to the above studies, which highlighted radially emitted THz radiation from plasma, suggests an alternative practical method for generating powerful longitudinal THz radiation from plasma filaments. D'Amico introduced a novel mechanism for generating THz emission in the air, offering the advantage of simplicity in 2007 [136]. In their experiment, a laser with an energy of 4 mJ per pulse, a repetition rate of 10 Hz, a central wavelength of 800 nm, and a width of 150 fs was focused in the air using a 2 m focal length lens to create a single filament. The spectral component of the THz radiation from the plasma filament was measured using a heterodyne detector. The experimental setup is shown in Fig. 9 [137]. This figure illustrates an experimental setup for generating and detecting THz waves using a laser-induced plasma filament. In this process, an ultrashort laser pulse is focused through a lens, leading to nonlinear ionization in the medium, which forms a plasma filament. Due to the interaction of the laser fields and nonlinear effects, the plasma filament emits THz waves. The emitted THz waves propagate in a conical pattern, characterized by the angles $\theta$ and $\alpha$. A metallic grid polarizer is used to filter and control the polarization of the THz waves before they enter a metallic waveguide. The waveguide directs the THz waves toward a detector for further analysis. This setup is commonly used in plasma-based THz generation studies, where the goal is to optimize THz wave efficiency and control their properties for applications in spectroscopy, imaging, and communication systems. Furthermore, an experimental setup for THz wave generation using plasma induced by a mid-infrared laser is depicted in Fig. 10. A femtosecond laser pulse (3.9 μm, 30 mJ, 95 fs, and 20 Hz) is initially polarized using a quarter-wave plate (QWP). It then passes through a 100 μm GaSe crystal, generating second-harmonic radiation ($2\omega$) alongside the fundamental frequency ($\omega$). The combined beams are focused by an off-axis parabolic mirror (OPM1) into a gas medium, creating a plasma filament. The plasma acts as a nonlinear medium, emitting THz radiation. The generated THz waves are collected and filtered using another off-axis parabolic mirror (OPM2) and a set of optical filters before being directed to a detector. The inset shows an image of the plasma filament, demonstrating its elongated structure and emission region. This setup is designed to optimize THz generation via plasma-based mechanisms, which have applications in spectroscopy, imaging, and ultrafast optics. The formation of the plasma filament results from the dynamic interplay between Kerr beam self-focusing and beam de-focusing caused by air ionization (Fig. 9). This competition leads to high peak intensity within a small average beam diameter (100 μm) over an extended distance. Throughout the filamentation process, the ponderomotive force of the laser field segregates charges at the pulse's peak intensity, leaving the plasma in an excited state of oscillations trailing the laser pulse. The primary oscillations induced by the transverse component of the laser field are transverse [138]. However, in the case of a femtosecond laser pulse, the net radial current is zero, resulting in no THz emission. Within the plasma filament, a slight longitudinal oscillation is also stimulated by the Lorentz force but is rapidly damped by electron collisions on a sub-picosecond timescale, ultimately responsible for THz emission. In this model, the source term of Lorentz force is derived from the study of Spangle et al [139]. The dipole-like localized plasma current density, resembling dipole moments moving at high velocities within the medium, emits radiation via the Cherenkov mechanism. When a charged particle passes through the interface between two media, such as when it traverses a boundary from a vacuum or one material to another, it experiences a sudden change in the local electric and magnetic fields. As a result, the particle emits electromagnetic radiation. This radiation spans a wide frequency range, including the THz regime. The characteristics of transition radiation in the THz regime depend on various factors, such as the energy of the charged particle, its velocity, the angle of incidence, and the properties of the media involved. The emission spectrum of transition radiation can be influenced by the refractive indices of the media, the angle of incidence, and the energy distribution of the incident particles. Nevertheless, as there is no discernible "transition" of particles occurring at the boundaries between the two media, it can be interpreted as a genuine blend of the two physical processes.

By investigating various aspects such as polarization, focusing conditions, Cherenkov angles, and plasma parameters, researchers have contributed to the growing body of knowledge surrounding the characteristics and applications of THz radiation in this respect. In a study conducted by Z. Wang et al., the focus was on the ultrafast imaging of THz transition-Cherenkov radiation in $LiNbO_3$ [141]. The researchers employed a technique whereby THz Cherenkov waves were generated through the optical rectification of laser pulses that were precisely focused into the $LiNbO_3$ material. The investigation in their work specifically explored the influence of polarization on the generation of this transition-like THz radiation. By analyzing the experimental setup and results, they were able to gain insights into the relationship between polarization and the characteristics of the generated THz radiation. In the study of Denis and colleagues, the generation mechanism of THz radiation from tightly focused femtosecond laser pulses in a gas medium by the numerical fit of the experimental data is conducted [142]. They measured the angular radiation pattern under different focusing conditions and observed that with the deepening of focus, the angular radiation pattern changes, and optical-to-THz conversion efficiency increases indicating the dominating role of dipole radiation in comparison to the quadrupole mechanism of radiation. Another group of researchers, A. P. Shkurinov et al., delved into the generation mechanism of THz radiation using tightly focused femtosecond laser pulses in a gas medium [143]. Through their experimental work, they examined the impact of dipole-like structures on the angular radiation patterns of THz waves under various focusing conditions.

Notably, they observed that as the focus deepened, the angular radiation pattern underwent significant changes. By studying these patterns, they were able to elucidate the intricate relationship between the focusing conditions and the resulting THz radiation characteristics. In a separate investigation, X. Lin et al. explored the effect of different Cherenkov angles on resonance transition radiation [144]. To accomplish this, they utilized photonic crystals to generate and manipulate Cherenkov radiation angles across a range of velocities, both in forward and backward directions. Through their experimental setup, they were able to observe and analyze the relationship between the Cherenkov angles and the resulting radiation properties. This research shed light on the interplay between the Cherenkov angles and the velocity-dependent characteristics of the generated radiation. Y. Chen and J. Zhao approached the study of THz radiation in both the time and frequency domains using the Cherenkov model [145]. They focused specifically on THz radiation driven by longitudinal current during single-color filamentation. Through theoretical analysis and experimental investigations, they probed the role of plasma density and electron collision frequency in determining the generated frequency of the THz radiation. Their work provided valuable insights into the fundamental factors influencing the spectral properties of THz radiation generated through the Cherenkov mechanism. In this respect, the spectral intensity of the transition-Cherenkov emission is given by [146]:

$$\frac{\partial^2 W}{\partial\omega\partial\Omega} = A_0 f(\theta) g(\omega, \alpha) \tag{7}$$

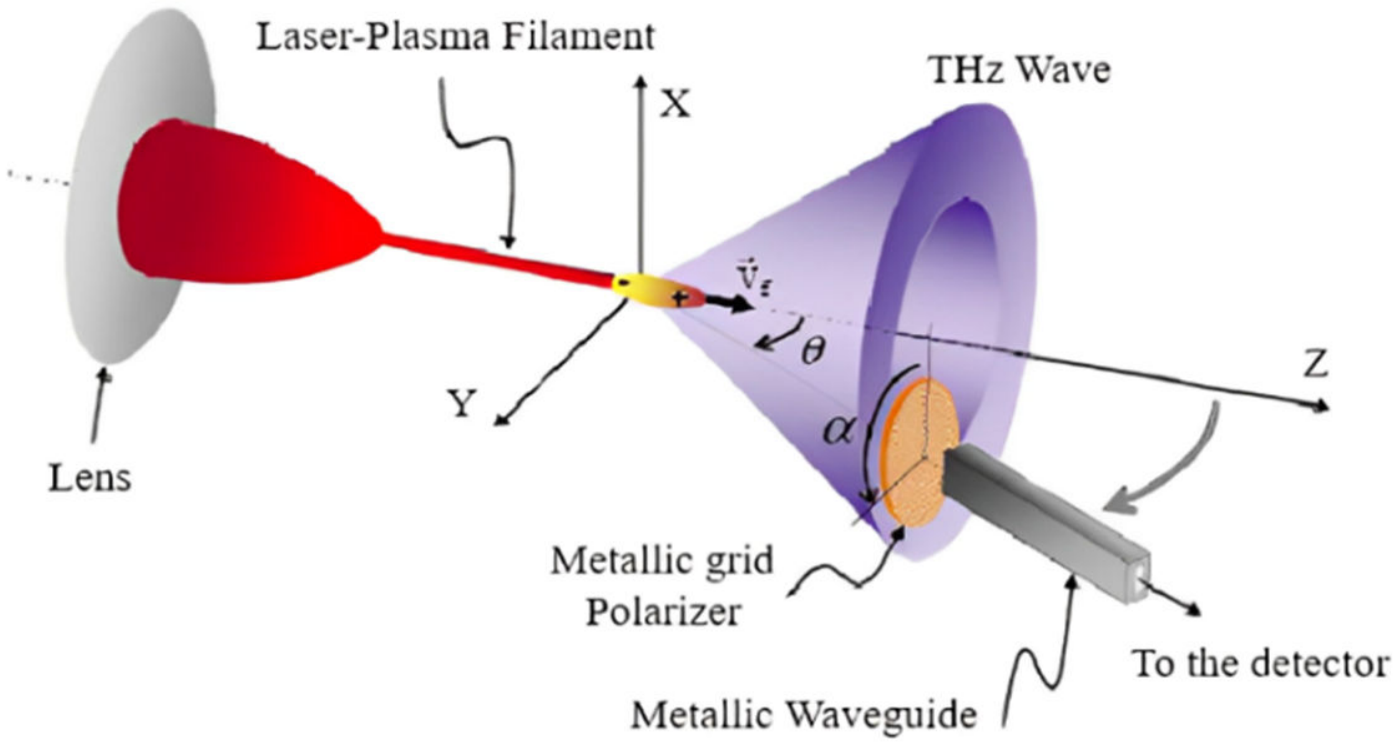

Figure 9: Experimental set-up used for THz generation and measurement [145].

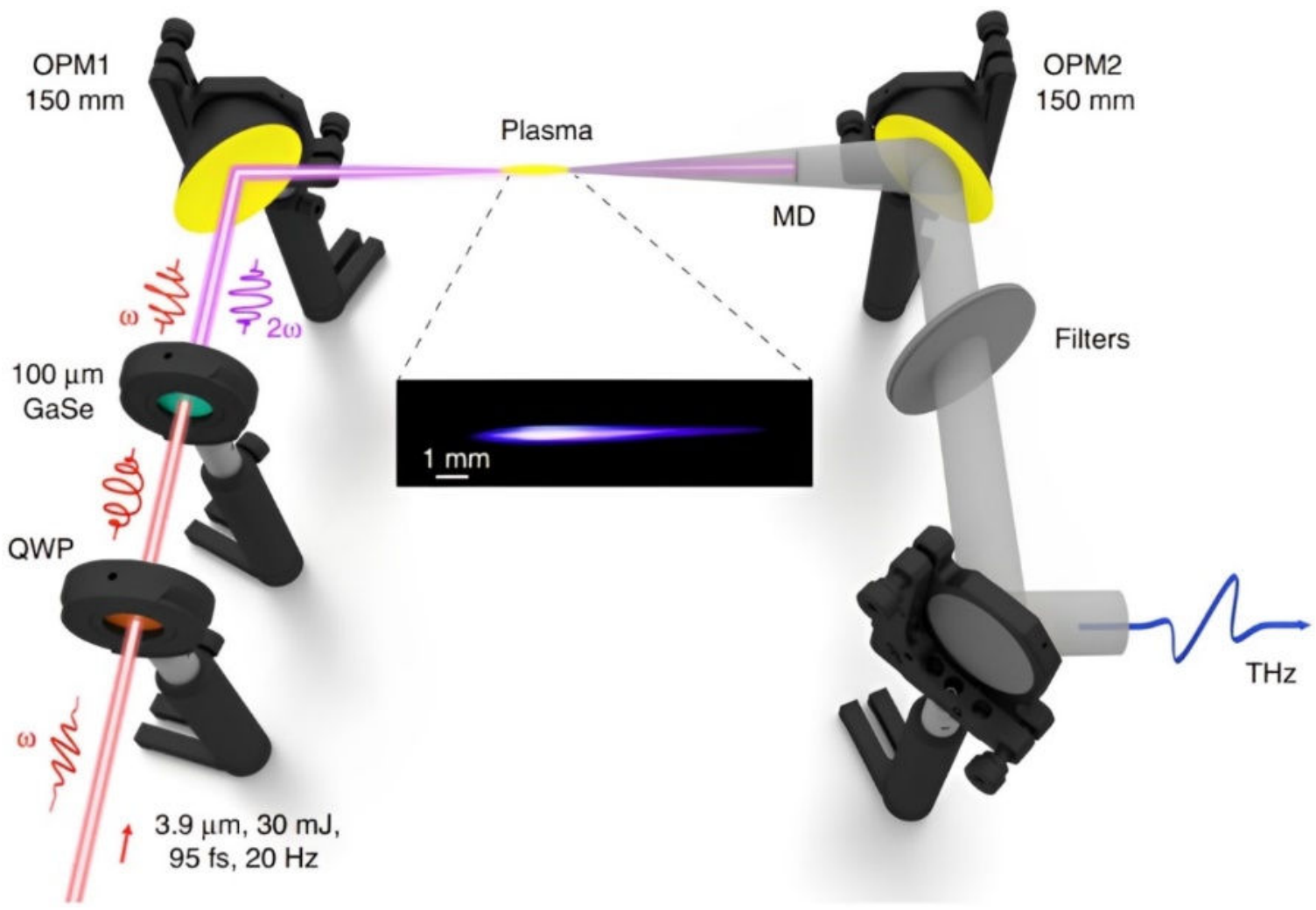

Figure 10: The configuration for THz generation via two-color mid-infrared filaments [140].

where $A_0 = \frac{\rho_0^2}{8\pi^2\varepsilon_0 c}$, the functions of $g(\omega,\alpha)$ and $f(\theta)$ are as follows:

$$g(\omega,\alpha) = \left[\frac{\varepsilon_0 E_{ext}\omega\omega_p^2}{\omega^2+i\nu_c\omega-\omega_p^2} + \frac{I_0\tau_L e\omega_p^2}{4m_e c}\frac{\omega+2i\nu_c}{\omega_p^2-\omega^2-i\nu_c\omega}\left(\frac{1+\alpha_1^2}{\nu_c^2+\omega_0^2} + \sum_{s\geq 2}\frac{1+\alpha_s^2}{\nu_c^2+s^2\omega_0^2}\right)\frac{Sin\left(\frac{\omega\tau_L}{2}\right)}{\left(\frac{\omega\tau_L}{2\pi}\right)^2-1}\right]^2 \quad (8)$$

$$f(\theta) = \frac{Sin^2\theta\left(e^{iL\left(\frac{\omega}{v_i}-kCos\theta\right)}-1\right)^2}{i\left(\frac{\omega}{v_i}-kCos\theta\right)^2} == \frac{\beta^2 Sin^2\theta Sin^2\left[\frac{L\omega}{2\gamma c}(1-\beta Cos\theta)\right]}{(1-\beta Cos\theta)^2} \quad (9)$$

where $L$ is the filament length, $v_i$ is the moving velocity of the dipole-like structures' localized current density equal to the speed of ionization front, and $\beta = v_i/c$. The function $f(\theta)$ is the conventional transition-Cherenkov radiation term, and the radiation spectrum in Eq. (9) is the conventional Cherenkov radiation modulated by the dipole-like structures' localized current density. The equation offers a succinct and clear depiction of the connection between the spectral intensity of transition-Cherenkov radiation and the velocity distribution function associated with dipole-like structures within a beam. This outcome suggests that the velocity distribution can be deduced from the spectrum of transition-Cherenkov radiation, and the resultant dipole-like current density moves at sub-luminous speeds. This trait holds significance as it differentiates the behavior of dipole-like structures and their accompanying radiation from other phenomena governed by the principles of Cherenkov radiation (Fig. 11).

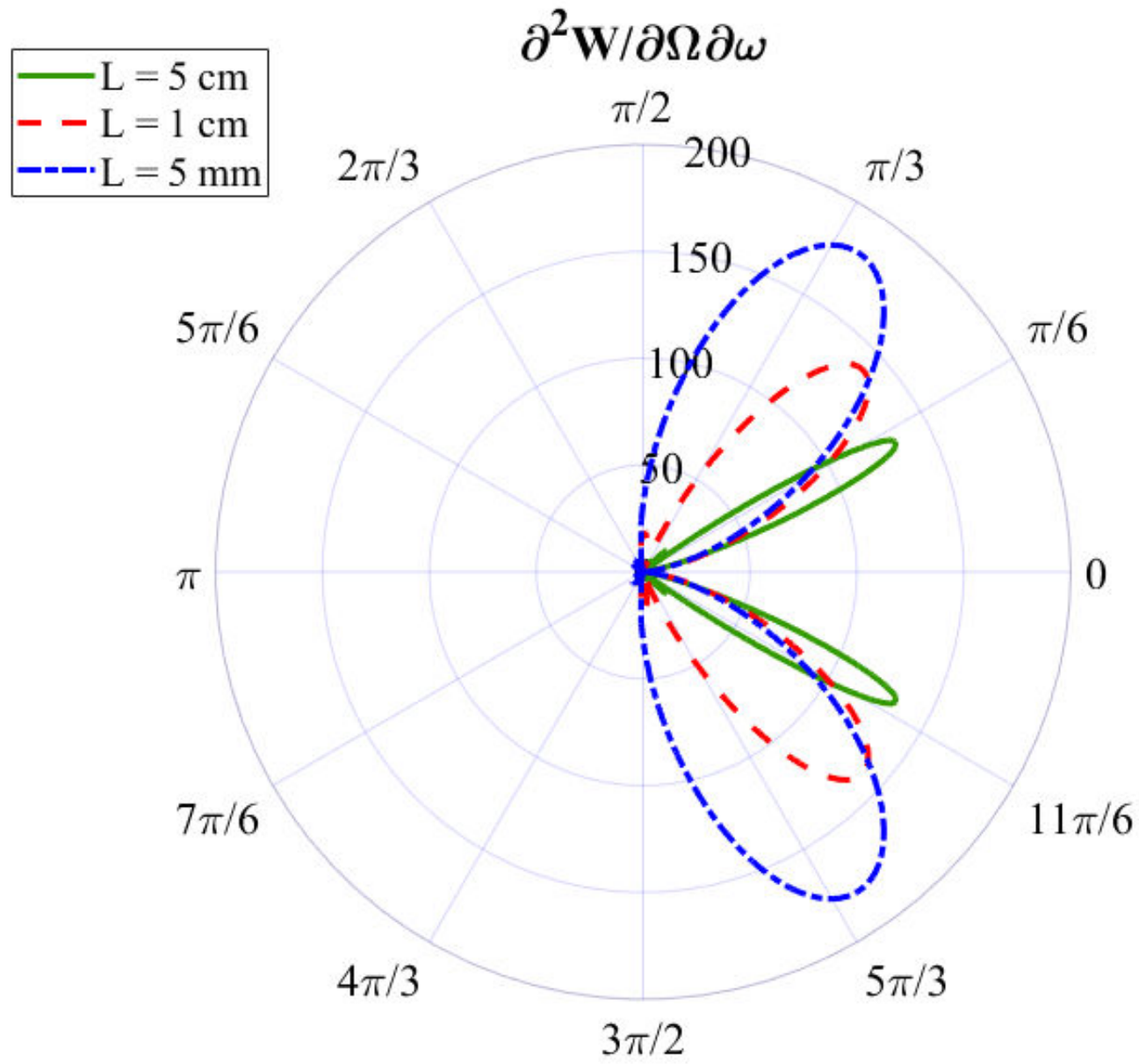

Figure 11: Variation of the angular distribution of THz radiation across various plasma lengths [138].

## 3.2 THz generation from laser-induced liquid-plasma

Generating THz waves using liquid water as a target has traditionally been deemed challenging due to its substantial THz band absorption. Liquid plasmas have emerged as promising broadband sources of THz waves due to their unique properties. Compared to gas media, liquid plasmas offer higher densities and lower critical potentials, making them suitable for generating intense THz radiation [147 - 149].
Since liquid plasmas have higher electron densities compared to gaseous plasmas, this increased density leads to enhanced THz wave generation through various mechanisms, such as coherent transition radiation and Cherenkov radiation. Moreover, liquid plasmas typically exhibit lower critical potentials, which allows for more efficient energy transfer and THz wave emission [150, 152]. Furthermore, the fluidity of liquids enables rapid replenishment of the plasma medium. This rapid replenishment is beneficial when the liquid is exposed to intense laser pulses for THz generation. It helps dissipate excess energy and prevents the material from reaching damage thresholds, thereby increasing the sustainability of intense THz wave generation. Liquid targets offer molecular equivalent cross-sections that are over three orders of magnitude greater than gas media and the probability of interactions between THz radiation and the liquid molecules is significantly higher, leading to more efficient THz wave generation. It is worth noting that the specific liquid used and the experimental setup play crucial roles in the performance of liquid plasma-based THz sources. Factors such as the selection of the liquid, its optical and thermal properties, and the laser characteristics need to be carefully considered and optimized to achieve the desired intense, broadband, and coherent THz radiation. In 2017, the experimental demonstration of broadband THz radiation from plasma within liquid water, induced by femtosecond laser pulses, marked a significant breakthrough. Subsequently, numerous experiments exploring different liquids and geometries have been conducted. Hereupon, comprehending the physical mechanism behind THz generation from liquid targets is crucial for the advancement of THz aqueous sources. The mechanisms of THz wave generation in liquid plasmas involve various physical processes that interact with the intense laser pulses. A dipole model, rooted in the ponderomotive force, is put forth to elucidate THz generation from laser-induced liquid plasma. Experimental results indicate that the THz energy in this direction is about 4.4 times that from the forward direction [153, 154]. CTR is a widely studied mechanism for THz wave generation in liquid plasmas. When an intense laser pulse interacts with a plasma, it drives the relativistic motion of the plasma electrons. As these electrons accelerate and decelerate, they emit electromagnetic radiation, including THz waves. The emitted THz radiation is coherent and exhibits a broad frequency spectrum. Cherenkov radiation occurs when a charged particle moves through a medium at a speed greater than the phase velocity of light in that medium. In liquid plasmas, the intense laser pulse induces the relativistic motion of electrons, which can result in Cherenkov radiation in the THz range. The emitted THz waves are coherent and directional, forming a cone-shaped radiation pattern.

A theoretical model that describes the THz generation from liquid water used the physics of ultrashort pulses and proposed by Tcypkin et al [155, 156]. In this plan the dynamics of the radiation field, the evolution of quasi-free electrons' current density, and the dynamics of electron concentration is considered. Based on three-coupled equations the variations of electric field and THz energy versus the pump pulse duration and liquid thickness was carried out as follows:

$$\frac{\partial E}{\partial z} + \Gamma_0 E - a_p \frac{\partial^3 E}{\partial \tau^3} + g E^2 \frac{\partial E}{\partial \tau} + \frac{2\pi}{c n_0} j = 0$$
$$\frac{\partial j}{\partial \tau} + \frac{j}{\tau_c} = \xi \rho E^3$$
$$\frac{\partial \rho}{\partial \tau} + \frac{\rho}{\tau_p} = \zeta E^2 \quad (10)$$

here $\Gamma_0$ and $a_p$ are the experimental coefficients that characterize the dependencies of the medium absorption and the refractive index, respectively. The parameter of $g$ describes the low-inertia cubic nonlinear response exhibited by the medium, $j$ is a current density composed of quasi-free plasma electrons, $\rho$ is population of excited states in the medium's molecules. The parameters of $\tau_c$ and $\tau_p$ are times of relaxation from quasi-free and excited states, respectively. The position of propagation is described by $z$, $\xi$ and $\zeta$ are the efficiency of the transitions to the above states. This model indicates the quasi-quadratic dependence of THz energy on the optical excitation. The maximum generation efficiency of THz radiation has been confirmed by experiments and simulations with this model which changes according to both the thickness of the liquid and the duration of the pump pulse.

The physical mechanism of THz generation by liquid plasmas can be explained by nonlinear optics and plasma physics. Simulation results are based on solving Maxwell's wave equation in paraxial approximation and consideration of three factors that correspond to the current density (third-order polarization, nonlinear absorption, and current density of free electrons along with the kinetic equation of concentration of free electrons), show that the model predictions are in good agreement with experimental results [157, 158]. The specific mechanisms of THz wave generation in liquid plasmas can depend on various factors, including the laser parameters, plasma density, and the properties of the liquid medium. The selection of the liquid and its optical properties can play a significant role in enhancing specific mechanisms and optimizing THz wave generation. When the laser pulse is focused on the interface between air and a water film, it penetrates and refracts through it, causing the ionization of water molecules at the focal point. In the ionized region, quasi-free electrons are propelled by the ponderomotive force toward regions with lower electron density, while the ions remain relatively stationary owing to their higher mass. The density of ionized carriers remains consistent in the forward direction since the envelope of the laser pulse advances faster than the electrons. As a result, the electrons are accelerated backward, creating a dipole structure that emits THz waves. The intensity of THz waves can be quantified as follows:

$$I_{THz}(\alpha, \beta) \propto T_1(\alpha) T_2(\beta) \sin^2[\gamma(\alpha, \beta)] exp\left[\left(-\frac{a_p d_f}{2\cos\theta_t(\beta)}\right)\left(\frac{W_L}{r_0}\right)^2\right] \quad (11)$$

where $\alpha$ is the incident angle of the laser pulse, $\beta$ is the detection angle of the THz wave, $T_1$ and $T_2$ are transmittance of the laser pulse and THz wave, $\gamma$ is the measurement angle related to the direction of the dipole, $\theta_t$ is the exiting angle of THz waves, $a_p$ is the absorption coefficient of THz waves, $W_L$ is the energy of the laser pulse, $d_f$ is the thickness of the water film. This Eq. indicates that the THz intensity from laser-induced liquid-plasma depends on the laser incident angle. Another mechanism responsible for the generation of THz radiation from a liquid water line involves a laser-induced ponderomotive force-induced current with broken symmetry around the water-column interface. In this model, THz radiation is believed to arise from a net current induced by the presence of the column interfaces. The currents distributed on both sides of the laser propagation axis are formed due to the ponderomotive force. If the laser axis deviates from the center of the water column, the symmetry of the two currents will be disrupted.

The choice of liquid medium depends on specific experimental requirements, such as desired THz frequency range, power levels, coherence, and compatibility with the laser system. Different liquids exhibit varying optical, thermal, and electrical properties, which can influence the efficiency and characteristics of THz wave generation [159]. Therefore, researchers carefully select and optimize the liquid medium based on their specific needs and the desired performance of the liquid plasma-based THz source. Several liquid mediums have been explored for THz wave generation in liquid plasma-based sources. Water has been explored as a medium for THz wave generation [151, 160], despite its relatively high optical absorption in the THz frequency range. Various mechanisms, such as laser-induced plasma in water, have been investigated to mitigate absorption losses and enhance THz emission efficiency. Water also possesses good thermal properties and is readily available, making it a convenient choice for experimental setups. Various organic solvents have been employed for THz wave generation. Examples include methanol, ethanol, isopropanol, and acetone. Organic solvents often exhibit low THz absorption and can offer favorable optical and thermal properties. The choice of solvent depends on factors such as compatibility with the laser system, solubility of desired substances, and specific experimental requirements [161]. Liquid nitrogen ($LN_2$) is used in cryogenic THz sources. By immersing the liquid plasma in $LN_2$, the temperature is significantly reduced, leading to lower thermal effects and improved stability. Cryogenic environments can enhance THz wave generation and reduce background noise. However, the use of $LN_2$ requires specialized cryogenic equipment and careful handling due to the extremely low temperature. Liquid helium ($LHe$) is utilized in extreme cryogenic THz sources. LHe offers even lower temperatures than $LN_2$, which can further reduce thermal effects and enhance THz generation. Cryogenic THz sources using LHe are typically complex and require advanced cryogenic systems and expertise. Liquid crystals have also been explored for THz wave generation. These materials possess unique optical properties, including birefringence and tunability, which can be harnessed for THz applications. Liquid crystals offer the advantage of active control over the THz wave properties, such as polarization and phase.

In this field, two independent experimental demonstrations of broadband THz radiation from plasma within liquid water and cuvette of liquids induced by femtosecond laser pulses, which were undertaken in 2017, marked a significant breakthrough in THz generation from liquid plasmas [162]. The THz emission from laser-induced filamentation in some liquids (methanol, ethanol, acetone, dichloroethane, deionized water, carbon disulfide) is reported. Free-standing water film is used in the experimental set-up where its thickness is chosen in a manner that strong absorption of water in the THz region does not affect the results [151, 163]. Experimental results indicate that the THz field from water film is 1.8 times stronger than that from air plasma. THz generation from laser filamentation in some liquids, when a single-color laser pulse is incident on liquids, indicated that the energy of THz emission is more than an order of magnitude higher than that obtained from the two-color filamentation of air plasma. In addition, numerous experiments exploring different liquids and geometries have been conducted to propose new methods for the generation of THz waves. Liquids with different polarity indices are used for THz emission. Results of this Ref. indicate that stronger THz fields can be generated from liquids with higher polarity indices. Liquid nitrogen and liquid gallium are other candidates that are used in experimental studies. A survey of the literature indicates that laser-induced plasma liquids are promising sources of intense THz waves in comparison to air plasmas.

### 3.3 THz generation from laser-induced clustered-plasma

As previously mentioned, intense THz radiation is highly sought after due to its wide range of applications and plasma has emerged as a promising medium for generating THz radiation by overcoming the issue of damage. However, increasing the pump laser intensity to further enhance THz yield has proven challenging. When the laser intensity reaches $10^{16} W/cm^2$, the ionization current saturates. Currently, ultra-intensity laser pulses can reach peak intensities above $10^{18} W/cm^2$, often referred to as relativistic lasers [164, 165]. At this level, the electron quiver velocity in the laser field approaches the speed of light.

Under these experimental conditions, it is desirable to have plasma without a saturated ionization current from the target. On the other way, compared to gases and liquids, nanosolids (clusters) can provide a higher density of electrons and withstand higher-intensity laser pulses. Presently, intense THz pulses with energy up to tens of mJ have been observed from cluster targets pumped by relativistic lasers [166]. There is no evidence of THz signal saturation even at pump intensities of $10^{20} W/cm^2$. Although obtaining intense THz waves through relativistic laser-cluster interaction is possible, the mechanisms involved are complex and still under exploration. The process of generating THz waves from laser-induced clustered plasma involves the interaction of intense laser pulses with clustered plasma, leading to the creation of THz waves. Clustered plasma refers to a state where atoms or molecules form distinct plasma structures by clustering together. Following rapid ionization (via MPI/tunneling, creating $n_e \approx 10^{21} cm^{-3}$) and heating ($T_e \approx U_p = (e^2E^2)/(4m_e\omega^2) \approx$ 10–100 eV), clusters transform into nanoplasma. Subsequent hydrodynamic expansion, driven by thermal pressure $P \approx n_e k_B T_e +$ radiation terms, forms plasma balls with radius growth dr/dt $\approx \sqrt{((\gamma P)/\rho)} \approx 10^5$–$10^6$ cm/s [165, 166]. When such clustered plasma is exposed to high-intensity laser pulses, nonlinear interactions occur, resulting in the emission of THz radiation. Studying the generation of THz waves from laser-induced clustered plasma is crucial for understanding the underlying physical processes and harnessing this phenomenon for various applications. Researchers are exploring this field to investigate various mechanisms, develop compact and efficient THz sources that can be used in spectroscopy, imaging, communication systems, and other applications.

The mechanism behind THz generation from laser-induced clustered plasma typically involves nonlinear processes through two-color laser fields. In this process, the intense laser pulses interact with the clustered plasma, initiating a stronger nonlinear current and a higher THz radiation output [167, 168]. The specific frequencies and characteristics of the emitted THz waves depend on the properties of the clustered plasma, the laser parameters, and the interaction dynamics. The atoms in clusters are rapidly ionized and transformed into plasma balls with overlying ion spheres and electrons when a fundamental harmonic femtosecond laser pulse combines with its second harmonic by a plano–convex lens and is co-focused on cluster elements. The electron cloud center begins to oscillate in all directions and is displaced by a distance relative to the ion sphere, this displacement produces a space-charge field. The electric fields of a two-color laser and space-charge electric field interact with clustered plasma, leading to the drift of plasma electrons and clusters along the electric field paths. As a result, an induced nonlinear current density is produced that is responsible for THz wave generation.

In the field of THz wave generation with nanosolids, several pieces of research are conducted to explore different aspects of THz emission from various states of matter and the potential of these substances as next-generation THz sources. Alexei V. Balakin et al., focused on the interaction of high-power femtosecond lasers with gas clusters [169]. Through experiments, they demonstrated the ability to differentiate between the contributions of different components to the THz signal generated by clustered and un-clustered plasma. Kazuaki Mori et al., studied directional linearly polarized THz emission from argon clusters irradiated by nonlinear double-pulse beams [170]. They identified the importance of appropriate time and optimized space intervals in the interaction between argon clusters and intense femtosecond double laser pulses. These parameters were found to enhance THz radiation generation, leading to power enhancement and high forward directivity. In other work, they explored the energy increase of THz waves from cluster plasma by controlling the laser pulse duration [171]. They showed that the energy of THz waves generated from an argon cluster plasma depends on the laser pulse duration and that by controlling the plasma density, THz energy can be increased. Rakhee Malik et al., investigated two-color laser-driven THz generation in clustered plasma [172]. They utilized the nonlinear mixing of the fundamental and second harmonic laser pulses in clustered plasma and demonstrated that THz radiation is enhanced through cluster plasmon resonance and phase matching introduced by density ripple. Ram Kishor Singh et al., conducted theoretical investigations on the generation of high-power THz radiation by beating two super-Gaussian laser beams co-propagating in clustered plasma [173]. Their study indicated that the cluster electrons experience a ponderomotive force at the beat wave frequency in both the axial and perpendicular directions to the beams' propagation. A. Frolov focused on the dipole mechanism for THz wave generation on a cluster under the influence of a laser pulse [174]. The study examined the dependence of the angular characteristics of THz emission on the electron density in the clustered plasma, particularly in conditions where dipole radiation predominates. It was found that the peak energy of THz radiation, known as the resonance condition, occurs when the laser frequency aligns with the natural oscillation frequency of a spherical cluster. Shivani Vij and Niti Kant analyzed the generation of THz radiation by beating two-color laser pulses in hot clustered plasma with a step density profile [175]. They presented a theoretical model for THz generation in clustered plasma and discussed the efficiency enhancement resulting from cluster plasmon resonance and the coupling between plasma and THz waves. Sharma and Vijay investigated resonant second-harmonic generation in magnetized plasma embedded with clusters [176]. They demonstrated that the insertion of clusters into magnetized plasma enhances the efficiency of second harmonic generation by intense laser pulses. Hai-Wei Du explored the role of second harmonic generation in broadband THz radiation. The study investigated the effect of the laser's second harmonic on THz generation through theoretical analysis and experimental processes [177]. E. Yiwen et al., conducted a review article investigating broadband THz emission from gases and liquids [178]. They examined the generation of THz waves using short laser pulse excitation and explored the properties and possibilities of these substances as THz sources. Arvind Kumar et al., investigated the plasma wave-assisted heating of collisional nano-cluster plasma through nonlinear interaction with two high-intensity lasers [179]. They showed that the anomalous heating rate is enhanced by surface plasmon oscillations resulting from the laser field interaction with embedded nano-cluster plasma. They also generalized this model to achieve soft X-ray emission via the Bremsstrahlung process. Collectively, these studies contribute to our understanding of THz wave generation and highlight the potential of clustered plasma, as sources for THz radiation.

### *3.3.1 Wave-mixing*

The four-wave mixing mechanism provides a versatile and efficient method for generating THz waves from femtosecond laser pulses interacting with clustered plasma. In this scheme, the fundamental and second harmonic of a laser pulse with frequencies $\omega_0$ and $2\omega_0$ propagating simultaneously in a clustered plasma, the density profile, and velocity vectors of the plasma electrons and clusters are modified [180]. This leads to inducing a nonlinear current density for THz wave generation and radiated electromagnetic fields. The electromagnetic fields of generated radiation affect the spectral and angular pattern of the THz emission. With the calculation of the radiated total energy, the total energy per unit, and solid angle per unit frequency can be evaluated as follows:

$$\frac{\partial^2 W_{THz}}{\partial\omega\partial\Omega} = \frac{c^2R^2}{2\pi} g(\theta,\varphi)|f(\theta,\omega)|^2 \tag{12}$$

where $g(\theta,\varphi)$ and $f(\theta,\omega)$ are:

$$g(\theta,\varphi) = J_{0x}^2(1 - sin^2\theta cos^2\varphi) + J_{0y}^2(1 - sin^2\theta sin^2\varphi) + J_{0z}^2 cos^2\theta \tag{13}$$

and

$$f(\theta,\omega) = \omega\tau_L e^{-\frac{\omega^2\tau_L^2}{4}} \frac{J_1\left(\frac{\omega r_0}{c}sin\theta\right)}{\frac{\omega r_0}{c}sin\theta} \left[\frac{sin\psi_1(\theta,\omega)L}{\psi_1(\theta,\omega)} + \frac{sin\psi_2(\theta,\omega)L}{\psi_2(\theta,\omega)}\right] \tag{14}$$

The Eq. (12) indicates that the direction and waveforms of THz radiation depend strongly on the laser-plasma parameters, such as laser wavelength, pulse duration, and plasma density.

### *3.3.2 Beating*

In the beating mechanism, when the plasma and clusters are exposed to laser pulses, they experience a phenomenon known as the beat frequency ponderomotive force. This force arises due to the combined effect of two laser pulses with slightly different frequencies interacting with the electrons in the clusters and is the difference between the frequencies of the two laser pulses. As the laser pulses propagate through the clustered plasma medium, they exert this ponderomotive force on the electrons within the clusters. The beat frequency ponderomotive force leads to a hydrodynamic pressure within the clusters, causing them to expand. This expansion transforms the clusters into plasma balls or plasma cores. The hydrodynamic pressure arises from the interaction between the laser pulses and the cluster electrons, causing them to oscillate in response to the beat frequency. This oscillation generates a pressure wave that expands the clusters. The interaction of the laser pulses with the plasma balls is a nonlinear process. In this case, the high-intensity laser pulses induce a nonlinear induced current density in the plasma balls. This induced current density is responsible for the generation of THz waves. Actually, laser-driven electron quiver ($v_q \approx e\,E/(m_e\,\omega) \approx 10^8$ cm/s) thermalizes via collisional absorption ($\nu_{ei} \approx 10^{12}s^{-1}Z^2 n_i Ln\Lambda/T_e^{3/2}$) and Mie resonances (enhancement factor $\approx (\omega_p/\omega)^2$ for spherical clusters), converting oscillatory energy to temperature ($\Delta T_e \approx v_{abs}I\tau/(n_e C_v)$, specific heat $C_v \approx 3/2k_B$). This generates internal pressure waves ($P \approx n_e k_B T_e \approx 10^8$–$10^{10}$ Pa), driving expansion and nonlinear currents for THz emission [173, 174]. The presence of cluster particles within the plasma balls plays a crucial role in enhancing the nonlinear induced current density and the efficiency of THz radiation generation. The cluster particles act as resonators, interacting strongly with the laser pulses. Their resonance with the laser frequency enhances the nonlinear response of the plasma, leading to a more significant generation of THz waves. Additionally, the behavior of the plasma in this scenario is influenced by several factors. Magnetic fields present in the plasma interact with the moving charged particles, altering their trajectories and affecting the oscillations of the electric field. Collisions between particles in the plasma also play a role, modifying the plasma density and temperature, further shaping the oscillations of the electric field. To assess the feasibility of clustered plasma as a source of THz radiation and explore efficiency enhancements, it is imperative to evaluate variations in the electric field and angular distribution of THz radiation. The electric field of the THz wave at the far-field region through the vector potential and the angular distribution of radiated power evaluated by the time-averaged Poynting vector leads to the quantity $\partial^2 W/\partial\omega\partial\Omega$ as follows [181]:

$$\frac{\partial^2 W}{\partial\omega\partial\Omega} = \frac{\mu_0 \tau_L c^2 r_P^4 (J_0^{THz})^2}{16\sqrt{\pi}}\left[1 - Sin^2\theta Sin^2\varphi\right] e^{-\frac{\omega^2\tau_L^2}{4}} \frac{Sin^2\left[\frac{L\omega}{2c}\left(1+\frac{3\omega p}{4\omega_1}-Cos\theta\right)\right]}{\left(1+\frac{3\omega p}{4\omega_1}-Cos\theta\right)^2} \tag{15}$$

This quantity can be utilized in the design of experiments for maximum radiation emission in specific directions and polarization. The plot shown in figure 12 illustrates the THz angular distribution for different cluster radii in both the beating and four-wave mixing mechanisms. The figure demonstrates that as the cluster sizes increase, the effect of boosted inner ionization becomes more prominent, leading to a higher induced nonlinear current density for THz emission. When compared to un-clustered plasma, the presence of clusters and the increase in cluster sizes contribute to a wider THz angular distribution and higher efficiency. The shape of the THz spectral response can be influenced by the size of the clusters. Smaller clusters tend to yield narrower spectral lines due to their limited degrees of freedom and simpler vibrational spectra. With fewer frequencies at which the clusters can resonate with the incident THz radiation, the resulting spectral lines are narrower. On the other hand, larger clusters lead to broader spectral features. This broadening effect arises from several factors, including the increased interaction time between the laser pulses and larger clusters. These prolonged interactions allow for more participation in collective motion effects and enhancement of resonance frequencies, resulting in a broader distribution of THz radiation. Additionally, the four-wave mixing mechanism generally generates THz waves with a wider bandwidth compared to the beating scheme. The nonlinear processes involved in four-wave mixing contribute to a broader range of frequencies being generated, resulting in a larger bandwidth for the THz waves produced. These findings provide valuable insights into the behavior of THz radiation generation in the presence of clustered plasma systems.

## 4 Discussion & Conclusion

This paper provides a comprehensive overview of THz radiation generation from laser-induced plasma in air, liquid, and clusters, focusing on the underlying mechanisms in each medium under various conditions. For air and liquid plasma, scenarios involving both relativistic and nonrelativistic pump laser pulses are considered, whereas for cluster plasma, the analysis is restricted to nonrelativistic laser intensities. The mechanisms are further categorized based on the characteristics of the pump laser fields, distinguishing between two-color and three-color interactions. Due to space constraints and the extensive literature, detailed discussions of relativistic interactions and their impact on THz energy yields from air plasma are not included. The properties of THz radiation from laser-induced plasma are examined individually for each medium, considering parameters such as THz energy and efficiency, spectral bandwidth, polarization, and angular distribution. As summarized in Table 2, THz generation from laser-induced gas plasma typically produces stronger signals than optical rectification in crystals, although the signal-to-noise ratio (SNR) is lower, highlighting the need for SNR enhancement. The spectral width of THz radiation from air plasma can extend up to ~100 THz, enabling ultrabroad absorption measurements, but detection techniques must be optimized to improve resolution and bandwidth. Two- and three-color schemes are widely used for laboratory THz generation from air plasma, yet their detailed mechanisms and operational limitations remain under investigation. Notably, THz waves from gas plasma often exhibit a hollow far-field distribution, which could be modified to a solid-core profile by reshaping the THz pulse or controlling the wavefront of the pump laser. In contrast, liquid plasma typically exhibits higher electron densities, leading to stronger THz signals under comparable pump conditions. However, THz field strength does not scale directly with electron density, likely due to absorption in the liquid. Enhancing THz generation in liquids requires increasing the plasma volume or length while minimizing absorption. Current theoretical models for liquid plasma are often adapted from gas plasma dynamics, underscoring the need for dedicated liquid-specific models. Although research on THz generation from liquids is still limited compared to gases and solids, strategies developed for gas plasmas—such as modifying target geometry or tuning pump laser wavelengths—can be applied to liquids to improve yield and bandwidth. THz radiation from laser-induced plasma also serves as a powerful diagnostic tool for studying plasma waves, electron transport, and particle acceleration. Beyond fundamental research, the THz band offers practical applications in communication technologies. As wireless communication evolves toward 6G, the THz spectrum is expected to enable data rates exceeding 1 Tbps, addressing challenges in holographic displays, multi-sensory communications, and other high-bandwidth applications that surpass the capabilities of current 5G networks. Laser–plasma THz sources excel in broadband spectroscopy, plasma diagnostics, and remote sensing (penetration through non-metallics ~ mm depths). While THz bands promise >1 Tbps data rates for 6G via ultra-wide bandwidth, plasma-based generation's high cost and complexity limit it to lab-scale or specialized secure links, unlike compact semiconductor alternatives [182].

Table 2: Comparison of THz sources [24, 39, 41, 147, 148].

| Mechanism | Power Output | Bandwidth | Efficiency | Applications | Limitations |
|---|---|---|---|---|---|
| **Laser-Plasma** | 10 nJ – 5 mJ | 0.1 – 100 THz | $10^{-5}$ – $10^{-3}$ | Spectroscopy, imaging, security | High complexity, lab-scale |
| **Difference Frequency Generation (DFG)** | 10 nJ – 10 µJ | 0.1 – 20 THz | $10^{-4}$ – $10^{-3}$ | Broadband spectroscopy | Crystal damage, phase-matching |
| **Photoconductive Antenna (PCA)** | 10 nJ – 1 µJ | 0.1 – 5 THz | $10^{-5}$ – $10^{-4}$ | Compact systems, THz-TDS | Low power, material breakdown |
| **Backward Wave Oscillator (BWO)** | 1 µJ – 1mJ | 0.1 – 1 THz | $3 – 10^{-2}$ | Continuous wave, communication | Vacuum requirements |

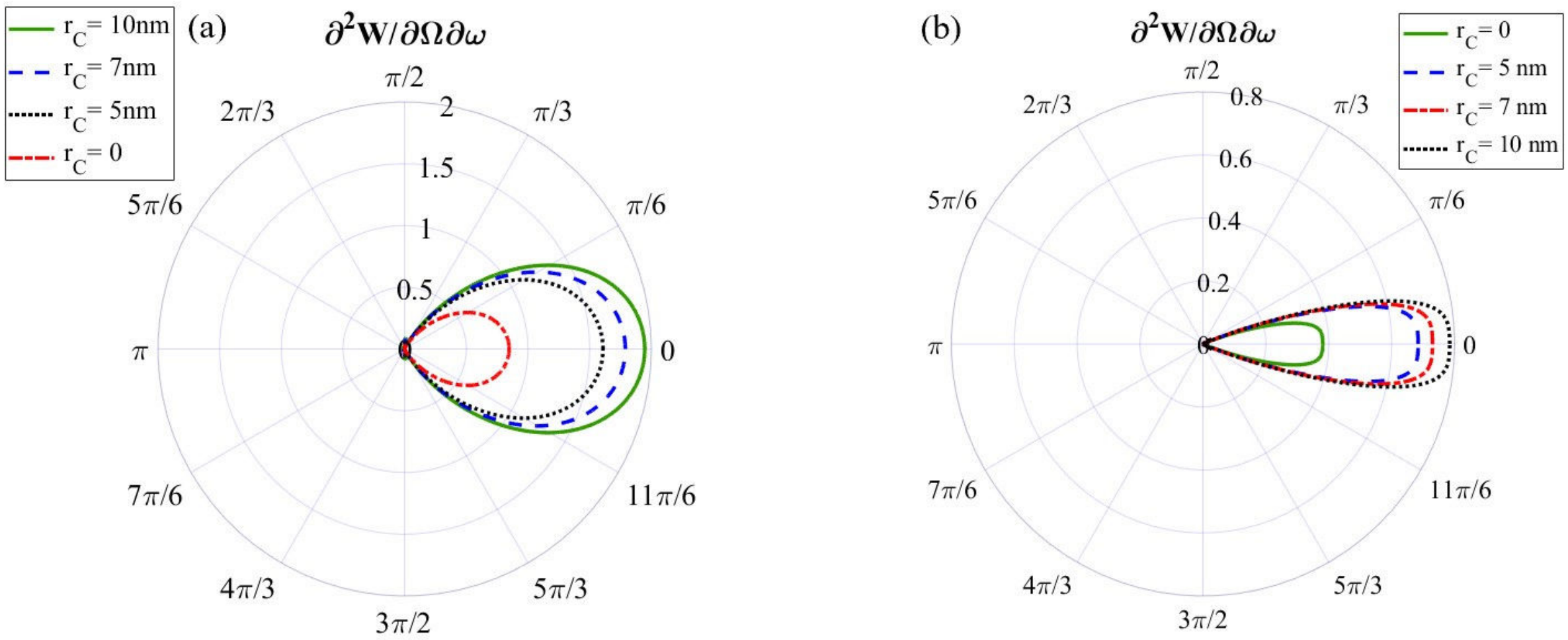

Figure 12: Variation of angular distribution of THz radiation for different values of cluster radii for (a) wave-mixing [198] and (b) beating mechanisms [199].

**Declarations, Competing interests:**
The authors declare that they have no known competing financial interests or personal relationships that could have appeared to influence the work reported in this paper.

**Data Availability:**
The datasets generated during and/or analysed during the current study are available from the corresponding author on reasonable request.

**Funding:**
This research did not receive any specific grant from funding agencies in the public, commercial, or not-for-profit sectors.